\documentclass[sigconf,nonacm]{acmart}
\AtBeginDocument{%
  \providecommand\BibTeX{{%
    Bib\TeX}}}
\usepackage{tabularx}
\newcolumntype{L}[1]{>{\raggedright\arraybackslash}p{#1}}
\newcolumntype{C}[1]{>{\centering\arraybackslash}p{#1}}
\newcolumntype{R}[1]{>{\raggedleft\arraybackslash}p{#1}}
\def\BibTeX{{\rm B\kern-.05em{\sc i\kern-.025em b}\kern-.08em
    T\kern-.1667em\lower.7ex\hbox{E}\kern-.125emX}}

\usepackage{tikz}

\usepackage{fontawesome5}
\usepackage{graphicx}
\usepackage{multirow}
\usepackage{bbding}
\usepackage{wasysym}
\usepackage{pifont}
\usepackage{soul}
\usepackage{ulem}
\usepackage[tableposition=top]{caption}
\usepackage{enumitem}
\usepackage{float}
\usepackage{amsmath}
\usepackage{tablefootnote}
\makeatletter
\newcounter{subsubsubsection}[subsubsection]
\usepackage{placeins}
\usepackage[most]{tcolorbox}
\newtcolorbox{insightbox}{
  colback=gray!10,    
  colframe=black,     
  boxrule=0.5pt,      
  arc=2mm,            
  left=2mm,           
  right=2mm,
  top=1mm,
  bottom=1mm,
  fonttitle=\bfseries,
  title=Insight:     
}

\newcommand\subsubsubsection{\@startsection{subsubsubsection}{4}{\z@}%
  {-1.5ex\@plus -1ex \@minus -.2ex}%
  {0.5ex \@plus .2ex}%
  {\normalfont\normalsize\bfseries}}
\newcommand\subsubsubsectionmark[1]{}
\makeatother

\def\l@subsubsubsection{\@tocline{4}{0pt}{1pc}{7pc}{}}

\begin{document}

\title{Advancing Security in Software-Defined Vehicles:  \\
A Comprehensive Survey and Taxonomy}

\author{Khaoula Sghaier}
\affiliation{%
  \institution{EPITA, VEDECOM}
  \country{France}
}
\email{khaoula.sghaier@epita.fr}

\author{Badis Hammi}
\affiliation{%
  \institution{SAMOVAR, Télécom SudParis, Institut Polytechnique de Paris}
  \country{France}
}
\email{badis.hammi@telecom-sudparis.eu}

\author{Ghada Gharbi}
\affiliation{%
  \institution{EPITA}
  \country{France}
}
\email{ghada.gharbi@epita.fr}

\author{Pierre Merdrignac}
\affiliation{%
  \institution{VEDECOM}
  \country{France}
}
\email{pierre.merdrignac@vedecom.fr}

\author{Pierre Parrend}
\affiliation{%
  \institution{EPITA}
  \country{France}
}
\email{pierre.parrend@epita.fr}

\author{Didier Verna}
\affiliation{%
  \institution{EPITA}
  \country{France}
}
\email{didier.verna@epita.fr}

\begin{abstract}
Software-Defined Vehicles (SDVs) introduce innovative features that extend the vehicle’s lifecycle through the integration of outsourced applications and continuous Over-The-Air (OTA) updates. This shift necessitates robust cybersecurity and system resilience. While research on Connected and Autonomous Vehicles (CAV) has been extensive, there is a lack of clarity in distinguishing SDVs from non-SDVs and a need to consolidate cybersecurity research. SDVs, with their extensive connectivity, have a broader attack surface. Besides, their software-centric nature introduces additional vulnerabilities. 
This paper provides a comprehensive examination of SDVs, detailing their ecosystem, enabling technologies, and the principal cyberattack entry points that arise from their architectural and operational characteristics. We also introduce a novel, layered taxonomy that maps concrete exploit techniques onto core SDV properties and attack paths, and use it to analyze representative studies and experimental approaches.
\end{abstract}



\maketitle

\section{Introduction}
\subsection{Scope of the paper}
\textit{Marc Andreessen} famously declared that “software is eating the world,” \footnote{\url{a16z.com/why-software-is-eating-the-world/}} and this statement rings particularly true today as software rapidly transforms and disrupts traditional industries. Nowhere is this shift more evident than in network-related technologies, where the transition from hardware-based to software-driven solutions is occurring at an unprecedented scale. Examples such as Software-Defined Networking (SDN) and Network Function Virtualization (NFV) vividly illustrate this paradigm shift, showcasing how software is revolutionizing the foundations of these industries. In this context, vehicles and Intelligent Transportation Systems (ITS) are no exception through the concept of Software Defined Vehicles (SDV) where software is emerging as the cornerstone of future automotive solutions fundamentally reshaping the transportation industry with initiatives like the AUTomotive Open System ARchitecture (AUTOSAR) which offers a standardized architecture for modular development, software reuse, and interoperability across embedded systems \cite{Autosar_standard}\cite{AUTOSAR_for_Connected_and_Autonomous_Vehicles_The_AUTOSAR_Adaptive_Platform}. 

The rapid expansion of SDVs necessitates increased attention to the critical area of cybersecurity vulnerabilities, both within vehicles (intra-vehicular) and in their external interactions (inter-vehicular). 
Recent strides in automotive software development, hardware abstraction, and the advent of novel vehicular use cases mandate an ongoing threat assessment and call for a diverse array of innovative security solutions to mitigate vulnerabilities. 
In recent years, numerous industry whitepapers and reports \cite{Deloitte}\cite{Bosch}, have sought to address the new demands and requirements of SDVs, delivering actionable insights for a safer and more intelligent SDV ecosystem. Nonetheless, addressing security concerns for SDVs remains significantly underdeveloped. The pervasive vehicular connectivity, combined with the elevated levels of driving autonomy, gives rise to highly targeted, covert, and scalable attacks that exploit the intrinsic vulnerabilities of intravehicular and extravehicular systems \cite{survey_SDV}\cite{article4}. For instance, a team of researchers at the University of South Carolina \cite{tesla_autopilot} successfully deceived the \textit{Tesla Model S'} autopilot system, causing it to perceive nonexistent objects or overlook real obstacles in the vehicle's path wirelessly and remotely. A further example of this is an internal Controller Area Network (CAN) bus attack through a \textit{JBL} speaker leading to the theft of a \textit{Toyota Model RAV4} in 2022 \cite{kentindell}. 
Furthermore, the growing use of Artificial Intelligence (AI) and Machine Learning (ML) in SDVs has enhanced existing functionalities and enabled a range of new applications. These technologies are applied in areas such as autonomous driving, including sensor fusion and object recognition, as well as in security measures like Intrusion Detection Systems (IDS) and Intrusion Prevention Systems (IPS). However, this increasing reliance on AI/ML can negatively impact the security and reliability of vehicle systems. 

Numerous studies \cite{On_data_fabrication_in_collaborative_vehicular_perception_Attacks_and_countermeasures}\cite{data_poison}\cite{data_poison_ITS} have examined the security of AI in Connected and Autonomous Vehicles (CAVs), highlighting distinct vulnerabilities across different phases. For example, \textit{Xiang et al.}  demonstrated powerful targeted attacks against 3D point-cloud classifiers, thereby compromising LiDAR-based perception \cite{Generating_3d_adversarial_point_clouds}. In parallel, sophisticated data poisoning strategies have been shown to corrupt the integrity of planning modules, allowing an attacker to manipulate vehicle trajectory prediction without explicit internal model knowledge \cite{Adversarial_backdoor_attack_by_naturalistic_data_poisoning_on_trajectory_prediction_in_autonomous_driving}.
Furthermore, state-of-the-art research has identified that evasion and poisoning attacks targeting intrusion detection systems can facilitate the misclassification of malicious traffic as benign and vice versa, thus circumventing established defense mechanisms within automotive networks \cite{Investigating_the_impact_of_evasion_attacks_against_automotive_intrusion_detection_systems}.

These aforementioned challenges motivate our investigation of cybersecurity in SDVs. Accordingly, we aim to present a comprehensive overview of the SDV ecosystem, covering both internal and external environments, along with practical real-world use cases. We analyze the cybersecurity threats and vulnerabilities, offering a comprehensive and detailed review of the current literature in this field. Besides, we propose a comprehensive taxonomy that systematically classifies SDVs-exclusive vulnerabilities across diverse domains.
\subsection{Related work}
CAVs and SDVs represent two key pillars for innovative mobility services. While CAVs focus on integrating connectivity and automation to enhance vehicle functionality, SDVs go a step further by using a software-defined architecture to enable greater flexibility, modularity, and integration with third-party solutions. 
However, this software-centric design makes SDVs more vulnerable to sophisticated and safety-critical cyberattacks due to their complex software environments and extensive interdependencies between multiple stakeholders.

Extensive research has been conducted in the field of e-mobility cybersecurity \cite{article2}\cite{article6}\cite{article10}\cite{article11}\cite{article13}\cite{article14}\cite{survey_SDV}, with considerable attention devoted to CAVs and Vehicle-to-Everything (V2X) communications. Existing surveys cover a broad spectrum, from attack taxonomies to mitigation techniques, and provide useful taxonomies and methodical reviews of defenses. 
Analyzing cybersecurity research in CAVs is relevant because SDVs are build upon similar foundational technologies. This perspective helps identify inherited threats, evaluate existing defenses, and provide a broader, more comprehensive view of cybersecurity challenges in SDVs.
In this context, \textit{Abdo et al.} \cite{article2} delved into various cyberattacks and defenses on CAVs, classifying attacks based on vehicle components such as sensors, communication channels, and internal networks, and exploring mitigation methodologies from cryptography to anomaly detection systems.

Considering connectivity as one of the most important features in CAVs, many surveys have studied the cybersecurity aspects of V2X \cite{article3}\cite{article4}\cite{article10}. 
For example, \textit{Sedar et al.} discussed in \cite{article3} the various attack scenarios and defense strategies specific to V2X communication. Their survey emphasizes the importance of secure communication protocols and the challenges in implementing robust security measures in a highly dynamic vehicular network. 
Despite its comprehensive focus on communication security, it lacks an extensive examination of in-vehicle systems and the interplay between internal and external threats. 
Moreover, while the study mentioned the multi-stakeholder ecosystem, the in-vehicle management of the multiparty security mechanisms was not evoked. 
In \cite{article4}, \textit{Suo et al.} survey security vulnerabilities and defense mechanisms for Vehicle-to-Vehicle (V2V) and Vehicle-to-Infrastructure (V2I) communications based on an integrated security engineering process at three levels : sensor level, vehicle level, and transportation level. However, there is a notable lack of focus on addressing attacks on critical vehicle-to-cloud attack surfaces that are essential for SDVs. 
\textit{Parkinson et al.} \cite{article10} have surveyed different vulnerabilities that a CAV would face from low-level sensors, behavioral aspects, and remote access perspectives. Although the authors have identified the cloud as a core CAV communication channel, they focus on threats to sensors, Electronic Control Units (ECUs), and V2V/V2I links, without providing any analysis of architecture or network-level vulnerabilities in vehicular cloud infrastructures. 
Furthermore, the authors highlight that aftermarket ECU programming modifications are often controversial, claiming that ECU software is proprietary by the Original Equipment Manufacturer (OEM). 
However, in the case of SDVs, there is a different approach; SDVs are designed to support, integrate and manage various third-party solutions for ECU software, offering a greater flexibility. 
To achieve this, an evolution towards white-box hardware, allowing for seamless integration of diverse software ecosystems using standards and openness, becomes inevitable. 
This transition mirrors trends in other software-defined domains, such as datacenters or networking, where white-box hardware has become a cornerstone for enabling innovation and reducing dependency on proprietary solutions. 
In this context, recent studies have increasingly focused on the automotive sector’s shift toward software-oriented architectures, examining the evolution of vehicles into SDVs \cite{article8}\cite{article5}. However, these works often do not fully engage with the security challenges inherent to the SDV ecosystem.

A key functionality of SDVs, and connected vehicles more broadly, is cooperative driving. This aspect has been studied by \textit{Amoozadeh et al.}  \cite{article12} from a security standpoint. Their study focuses on vulnerabilities affecting vehicles' components such as sensors and technologies like wireless V2X communication that could compromise the security and thus the safety of the cooperative driving vehicles within V2V and V2I communication. 
However, a limited range of use cases related to the vehicle connectivity were explored. Besides, attacks on road infrastructure components were not addressed, despite their relevance in the V2X communication. In \cite{article11}, \textit{Petit et al.} have examined the vulnerabilities from both autonomous vehicles and cooperative transportation perspectives. 
Their study elucidates the distinctions between these two paradigms, underscoring the critical necessity for implementing robust security countermeasures to mitigate the identified risks. Nonetheless, their focus was exclusively on autonomous and cooperated automated vehicles, leaving out the broader concept of SDV. 

\begin{table*}[t!]
\begin{center}
\footnotesize
\begin{tabular}{|L{2.1cm}||C{0.5cm}|C{5.2cm}|C{5.2cm}|C{1.3cm}|}
    \hline
         Paper & Year & Description & Limitations & SDVs coverage\\
         \hline
         \hline
         \textit{Petit et al.} \cite{article11} & 2014  & Cyberattacks on automated vehicles, attack vectors and their implications & OTA updates and vehicular cloud cyberattacks were not covered  & \Circle\\
         \hline
         \textit{Amoozadeh et al.} \cite{article12}  & 2015  & Security vulnerabilities of connected vehicle streams and their impact on cooperative driving &  Attacks on vehicular infrastructure (e.g. RSUs, edge, cloud) were not addressed & \Circle \\
         \hline
         \textit{Parkinson et al.} \cite{article10} & 2017 & CAV cyber threats: internal sensors, external communication channels, and behavioral aspects & Partial SDV attack-surface coverage. Ecosystem interactions underexplored & \LEFTcircle \\
         \hline    
        \textit{Vdovic et al.} \cite{article8}  & 2019  & Software aspects of CAV and electric vehicles, current technologies, challenges, and future directions & Lack of detailed discussion on automotive security aspects & \LEFTcircle \\
         \hline
         \textit{Elkhail et al.} \cite{article14} & 2021 & Security issues vulnerabilities, in-vehicle malware attacks, and attack detection mechanisms & Lack of comprehensive coverage of security concerns on multi-supplier solution management & \LEFTcircle\\
         \hline
         \textit{Suo et al.} \cite{article4} & 2022  & CAV security vulnerabilities and countermeasures from an engineering design perspective & V2X wireless communication technologies security analysis were not provided & \Circle\\ 
         \hline             
         \textit{Abdo et al.} \cite{article2} & 2023 &  Threat analysis and risk assessment of CAV vulnerabilities and countermeasures & Focus limited to a specific range of sensors and application-based attacks & \Circle \\ 
         \hline
         \textit{Sedar et al.} \cite{article3} &  2023 & V2X threats and security mechanisms with a proposed taxonomy & In-vehicle CAV attacks were not addressed & \Circle\\ 
         \hline
         \textit{Blanco et al.} \cite{article5}  & 2023  & Transformation of automotive systems towards SaaS models, benefits and challenges & Coverage limited to a subset of automotive software components & \LEFTcircle\\ 
         \hline
         \textit{Hossain et al.} \cite{article13} & 2023  & Categorization and analysis of CAV security attacks & Insufficient attention to supply-chain threats, such as compromises of third-party software vendors or dependencies & \Circle \\
         \hline
         \textit{De Vincenzi et al.} \cite{survey_SDV} & 2024 & Security and privacy concerns in SDVs: cyberattacks and countermeasures &  Lacks systematic coverage of key dimensions and detailed attack descriptions framed by explicit threat models & \CIRCLE \\
         \hline
    \end{tabular}
    \caption{Comparison of recent existing surveys on security in the automotive domain. \Circle: No; \LEFTcircle: Briefly; \CIRCLE: Yes.} 
    \label{table:surveys} 
\end{center}
\end{table*}

\textit{Hossain et al.} \cite{article13} have discussed various types of cyber-attacks on CAVs, including those targeting vehicular networks, sensor vulnerabilities, hardware exploits, and adversarial attacks. While the authors highlighted the variety of OEMs and tier-one suppliers involved in vehicular systems, they did not address the critical cybersecurity concerns within the supply chain. Additionally, potential vulnerabilities arising from software partitioning and inter-communication between the virtualized environments within the vehicle were notably absent from their analysis. 

\textit{Elkhail et al.} \cite{article14} have detailed the architecture of intelligent vehicles, encompassing ECUs, in-vehicle network, and sensors to further introduce the attack vectors and entry points. However, although the authors examined software update mechanisms, they did not extend their analysis to the in-vehicle management of software from multiple stakeholders, a critical consideration for SDVs. 
\textit{De Vincenzi et al.} \cite{survey_SDV} offer a structured literature review and expert elicitation that map security vulnerabilities relevant to SDVs. Their treatment addresses important topics like insecure Application Programming Interfaces (APIs), Over The Air (OTA) update channels, and in-vehicle networks, but it omits several systemic attack classes such as lateral movement and privilege escalation, where an adversary initially compromises low-privilege components (e.g., infotainment) and then escalates access to safety-critical functions (e.g., collision-avoidance systems) in mixed-criticality Service-Oriented Architectures. Moreover, although the authors acknowledge prior work on sensor attacks in the CAV literature, certain adversarial vectors become especially salient at higher autonomy levels (Level 3+), because SDVs increasingly depend on multi-modal sensor fusion. This dependency changes both the attack surface and the failure modes, and thus deserves more focused analysis.

Although CAVs and SDVs are integral components of the broader intelligent transportation ecosystem, sharing key attributes such as connectivity, automation, and the need for robust cybersecurity measures, there are crucial distinctions that render SDVs more vulnerable to severe and safety-critical attacks. 
This heightened vulnerability is due to their more complex cybersecurity landscape and exposure to an extensive software environment that includes a continuous evolution and upgradability. Although some surveys have addressed software vulnerabilities and challenges linked to OTA updates \cite{article8}\cite{article10}\cite{article14}, the intricacies of OTA updates in SDVs extend far beyond what has been previously explored. 
These updates encompass a network of interconnected systems and devices, alongside a diverse array of stakeholders actively participating in vehicle configuration and management processes. 
These third parties maintain continuous communication with the OEM to operate the OTA updates, upgrade various devices such as ECUs, and reorganize the in-vehicle communication data. Additionally, advanced features, such as automotive computing in the cloud \cite{cloud_qnx}, have not been adequately addressed in existing literature.

Table \ref{table:surveys} summarizes key surveys, reviews, and Systematization-of-Knowledge (SoK) papers on automotive cybersecurity, with concise notes on each work’s scope and limitations. Among these references, only \cite{article8} explicitly uses the term Software-Defined Vehicle, yet it merely signals the software-centric architectural shift and does not address SDV-specific features or security considerations. 
Note that this section is deliberately limited to surveys/SoKs and does not include works that target a single subsystem (e.g., \textit{Namburi et al.} \cite{Software_Defined_Vehicle_Fleet_Management_System_with_Integrated_Cybersecurity_Measures}, which focuses on fleet-management security). 
More recently, \textit{De Vincenzi et al.} \cite{survey_SDV} provided a structured review mapping SDVs vulnerabilities and mitigations. Nevertheless, the study lacks systematic coverage of key dimensions and detailed attack descriptions framed by explicit threat models. 
Accordingly, our paper is well positioned to fill these gaps by offering a systematic, up-to-date overview of SDVs threats that spans both in-vehicle components and interactions with the wider transportation infrastructure, thereby enabling a more holistic assessment of SDVs security.

\subsection{Contributions of this work}
This work aims to offer researchers from both industry and academia a thorough understanding of cybersecurity challenges within the SDV ecosystem. We highlight the contributions of this paper as follows:
\begin{enumerate}
    \item We provide an overview of the SDV ecosystem, encompassing both internal and external aspects, and highlight its distinctions from the CAV framework regarding applications, deployed architectures and enabling technologies. 
    \item We provide an in-depth examination of the internal network architectures within SDVs, including the roles of zonal ECUs, the implementation of the Automotive Ethernet (AE) protocol, and the integration of advanced telematics systems. 
    \item We assess the exposure of the vehicle’s High-Performance Computer (HPC), the onboard computing platform that handles SDV functions, to external connectivity technologies (e.g., Cellular V2X and 5G) accessed via the telematics unit, and discuss the resulting security implications.
    \item Building on these architectural and connectivity insights, we systematically identify and categorize the security threats and vulnerabilities specific to the SDV ecosystem, and organize them into a novel, comprehensive taxonomy.
\end{enumerate}

The remainder of this manuscript is structured as follows. Section \ref{sec:background} introduces the key characteristics of SDVs, describes the transition from current vehicle architectures to SDV architectures, highlighting their differences and advantages, and surveys the enabling technologies. Section \ref{sec:security} defines the threat model that underpins our analysis and then presents the proposed taxonomy. Section \ref{sec:summary} discusses existing studies on cybersecurity in SDVs, using our taxonomy as a framework to highlight remaining challenges. Section \ref{sec:challenges} addresses these challenges and open issues, proposing strategies and methodologies to overcome them, thereby laying the groundwork for secure-by-design systems aligned with cybersecurity standards. Finally, Section \ref{sec:conclusion} concludes the manuscript by summarizing key findings and outlining future research directions.

\section{Background} 
\label{sec:background}

This section provides background information on key concepts and terminology related to SDVs.

\subsection{SDVs: definition and architecture}
\label{sdv_def}
SDV represent a paradigm shift where software dictates vehicle functionality, surpassing traditional hardware dependencies \cite{article1}.
Hence, the SDVs are characterized by four key attributes:
\begin{enumerate}
    \item Flexibility and adaptability: new features are introduced via software updates, minimizing hardware modifications and ensuring technological relevance.
    \item Autonomous driving capabilities: SDVs integrate Advanced Driver Assistance Systems (ADAS), machine learning capabilities, and sensor data to achieve various autonomy levels, surpassing non-SDVs in AI and software-driven decision-making. 
    \item Improved user experience: personalized infotainment systems and seamless mobile integration make the driving experience more enjoyable and convenient\footnote{For instance, SDVs can leverage AI-powered hyper-personalization to automatically adjust seat positions, climate control, lighting, and music playlists based on individual driver preferences, biometric data, or even the time of day and destination. Additionally, SDVs enable subscription-based feature activation, allowing drivers to temporarily unlock advanced functionalities such as enhanced performance modes for towing, premium lighting systems for rural driving, or specialized ADAS features on-demand without requiring hardware modifications \cite{Experience_meets_engineering_infotainment_in_the_software_defined_vehicle_SDV}\cite{What_Is_a_Software_Defined_Vehicle}.}.
    \item Enhanced connectivity: V2X communication capabilities enhance safety, traffic management, and operational efficiency.
\end{enumerate}

To support continuous updates and enhancements, the Electrical and Electronic (E/E) architecture underlying today’s vehicles requires substantial revision, as the prevailing distributed model is increasingly unable to meet the computational, connectivity, and scalability demands of software-defined functionality.
In traditional distributed setups, each ECU is dedicated to a specific function, tightly integrating hardware and software. This one-to-one model leads to repetitive software development and inefficient resource use \cite{Deloitte}. For example, in non-SDVs, ECUs independently process shared sensor data, resulting in redundancy.
SDVs optimize this approach through zonal distribution, which consolidates functionality into fewer, high-performing zonal ECUs \cite{cesa}. These zonal ECUs directly control sensors and share data with one another and the High-Performance Computer (HPC) to coordinate actions \cite{Jama}. This architecture minimizes redundancy and enhances efficiency.

\begin{figure}[h]
    \centering
    \includegraphics[width=1.03\linewidth]{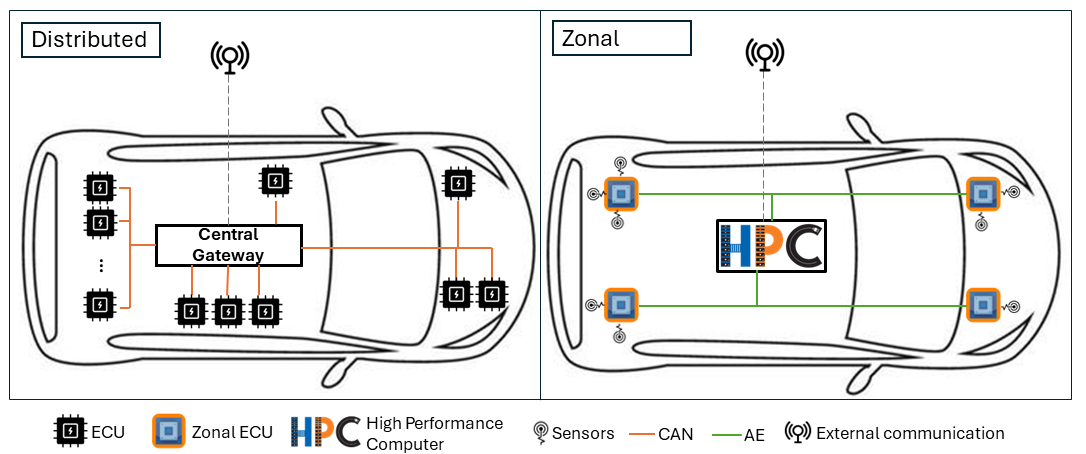}
    \caption{Electrical and Electronic (E/E) architecture: Distributed vs Zonal}
    \label{fig:architecture}
\end{figure}

Figure \ref{fig:architecture} illustrates the architectural transition in SDVs from a traditional distributed model to a zonal architecture. 
In conventional distributed vehicle architectures, each ECU is dedicated to a specific function, with hardware and software tightly coupled. While this one-to-one model ensures functional separation, it leads to repetitive software development, inefficient use of resources, and redundant sensor data processing across multiple ECUs \cite{Deloitte}. 
Moreover, domain-specific ECUs are interconnected through CAN networks and routed via a central gateway, which increases wiring complexity, limits scalability, and hinders centralized implementation of security measures or efficient OTA updates. To overcome these limitations, SDVs adopt a zonal architecture that consolidates functionality into fewer, high-performance zonal controllers \cite{cesa}. 
These zonal controllers directly interface with sensors and actuators grouped by physical location, while sharing data both horizontally across zones and vertically with a central HPC. 
The HPC serves as the system’s computational core, coordinating vehicle-wide functions and enabling advanced features \cite{Jama}\cite{renault}.
Furthermore, the zonal architecture significantly reduces wiring, improves data flow efficiency, and enhances scalability. It also supports OTA updates for deploying new features, such as adaptive cruise control, lane-keeping assistance, or personalized infotainment settings, without compromising the system's performance. 
The HPC’s direct interaction with the vehicle’s Operating System (OS) allows for isolation between software domains, secure boot processes, and runtime integrity checks, thereby maintaining a robust and secure software environment. 
Also, the Telematics Control Unit (TCU) plays a critical role by linking the HPC to the OEM cloud and enabling V2X communication. This supports real-time data exchange with infrastructure and other vehicles, improving both safety and traffic efficiency. 
Overall, this architectural evolution enhances performance, security, and the seamless integration of internal and external vehicular systems \cite{article1}.

\begin{figure}
    \centering
    \includegraphics[width=1.1\linewidth]{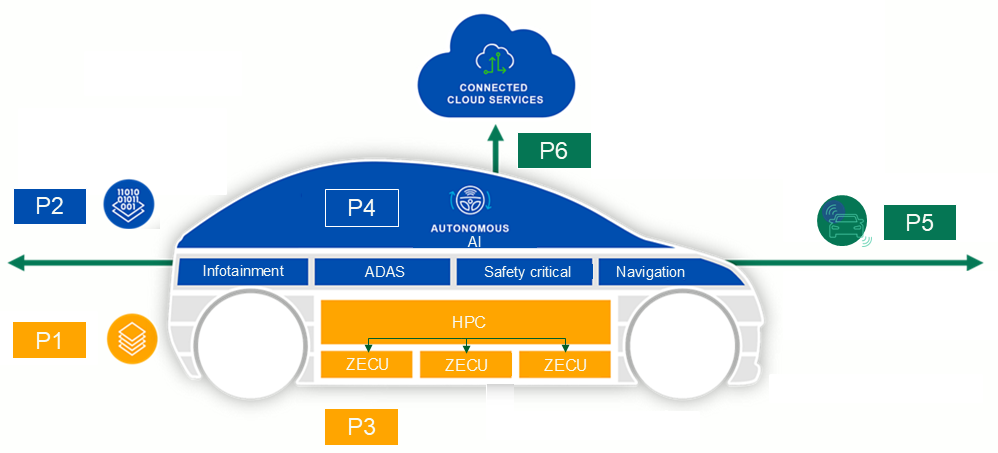}
    \caption{SDV relative properties. \textit{Color key:} Orange: Architectural (core on-board design and compute topology). Blue: Connectivity (internal/external dataflows, OTA, telematics, V2X). Green: Ecosystem (supply chain, cloud computing, third-party components, lifecycle/maintenance). \textit{P-labels:} P1: Hardware–software decoupling; P2: Modularity; P3: The zonal architecture; P4: Reliance on AI, P5: V2X and connectivity; P6: Supply chain.}
    \label{fig:sdv_properties}
\end{figure}
The shift to centralized processing in SDVs enables advanced driving tasks and real-time data processing but also increases communication bandwidth requirements and expands the system attack surface. This broader attack surface requires a systematic alignment of threat scenarios with the defining properties of SDVs. Accordingly, we selected our attack vectors by mapping them directly to the core characteristics of SDVs. As Figure 2 illustrates, these principal properties (P1 to P6) are summarized as follows:

\begin{itemize}
    \item \textbf{P1 - Hardware-software decoupling:} a fundamental characteristic of SDVs is the decoupling of software services from the underlying hardware, achieved through abstraction layers and virtualization techniques. This separation enhances software modularity, facilitates OTA updates, and allows services to run across different hardware platforms without modification. However, this architectural shift also broadens the attack surface. Since critical services no longer depend on specific hardware components, attackers can target vulnerabilities in shared software layers, middleware, or APIs, potentially compromising multiple functionalities at once (denoted by \faCloud{} and \faBell{}\footnote{Each pictogram denotes an attack category. These categories are defined in the "Attack" column of the taxonomy and further explained in subsection \ref{taxonomy}}). The absence of strict hardware binding increases the risk of privilege escalation, lateral movement, and service spoofing, particularly in cloud-connected and notification-based components.
    
    \item \textbf{P2 - Modularity:} the separation of software layers in SDVs enables modularity and composability, allowing functionalities to be organized based on mixed-criticality. Safety-critical and non-safety-critical functions can be isolated into distinct partitions, each governed by tailored access control mechanisms and security protocols according to its criticality level. However, this modular architecture also introduces attack vectors, as adversaries may attempt unauthorized access to escalate privileges and compromise restricted areas (denoted by \faWindowRestore{} and \faBuromobelexperte{}).
    
    \item \textbf{P3 - The zonal architecture:} the evolution of E/E architecture has led to the adoption of new protocols, such as SOME/IP, to support its service-oriented nature. While SOME/ IP facilitates seamless communication between software services, it lacks built-in security mechanisms, making it a potential attack vector for cybercriminals seeking unauthorized access to the in-vehicle network (denoted by \faMicrochip{} and \faInfinity{}).
    
    \item \textbf{P4 - Reliance on AI:} SDVs start at SAE automation level 3\footnote{SAE J3016 defines six automation levels (0–5). 0: No automation. 1: Driver assistance. 2: Partial automation (system controls steering and/or speed while the driver monitors). 3: Conditional automation (system drives within a defined Operational Design Domain (ODD) but may request human takeover). 4: High automation (system performs and monitors driving within its ODD without human intervention). 5: Full automation (system handles all driving in all conditions)} \cite{SAE}. Therefore, they deeply integrate AI into perception, decision-making, and control systems. However, this reliance extends the attack surface beyond the AI models themselves, encompassing the entire data and inference pipeline \cite{Exploring_Software-Defined_Vehicles_A_Comparative_Analysis_of_AI_and_ML_Models_for_Enhanced_Autonomy_and_Performance}. Adversaries can exploit vulnerabilities at multiple levels, including sensor inputs (e.g., spoofing, jamming, adversarial perturbations), data fusion and preprocessing stages, and even the deployment environment (e.g., compromised runtimes or exposed APIs). These attacks can manipulate AI behavior without altering the model directly, leading to misclassification, unsafe decisions, or loss of control. Moreover, poisoning attacks or the insertion of backdoors during training pose persistent threats, especially in systems relying on third-party datasets or pre-trained models.  This heightened risk underscores the need for robust security measures to safeguard AI stack against adversarial threats (denoted by \faDoorOpen{} and \faMapSigns{}).
    
    \item \textbf{P5 - V2X and connectivity to the external world:} SDVs are deeply integrated into the connected ecosystem, relying on V2X and C-V2X communications to interact with infrastructure, cloud services, and other vehicles. This extensive connectivity enables advanced functionalities, such as real-time traffic coordination and cooperative driving, but also significantly broadens the attack surface. Adversaries can exploit vulnerabilities in 5G and beyond networks, manipulate Quality of Service (QoS) parameters in V2X, or intercept critical messages to disrupt vehicle operations, making the security of these communication channels a major concern (denoted by \faBroadcastTower{}, \faHands{}, and \faSignal{}).
    
    \item \textbf{P6 - Supply chain:} the development and operation of SDVs depend on a complex, global supply chain involving multiple stakeholders, from hardware manufacturers to software vendors and cloud service providers. While this ecosystem enables rapid innovation and scalability, it also introduces critical security risks. Compromises at any stage, such as firmware tampering or software supply chain attacks, can have cascading effects on the entire SDV infrastructure. Attackers may exploit vulnerabilities in third-party components to inject malicious code, manipulate updates, or disrupt critical vehicle functions, making supply chain security a paramount concern (denoted by \faPuzzlePiece{} and \faSitemap{}).
\end{itemize}
\begin{table*}[!htbp]
\footnotesize
\begin{center}
\begin{tabular}{|L{1.9cm}||C{1.9cm}|C{6.1cm}|C{6.1cm}|}
    \hline
    Application & Feature & Non-SDVs & SDVs \\
    \hline
\hline
    \multirow{3}{2.1cm}{ADAS} & Hardware complexity & One-to-one matching model (one ECU, one task) & Higher level of hardware abstraction \\ \cline{2-4}
    & Sensor data handling & Sensor data is processed locally within individual ECUs & Various sensor data are processed centrally, managed and interpreted by software \\ \cline{2-4}
    & Lifecycle management & Limited; upgrades require physical changes to hardware components & Highly flexible; updates and continuous software improvements through OTA updates \\
    \hline
    \multirow{4}{2.1cm}{Connected car services} & V2X services & Limited; often require hardware add-ons for such functions & Extensive; integrated into larger mobility ecosystems, enabling features like ride-sharing, automatic electric vehicle charging, and seamless interaction with smart cities through V2X \\ \cline{2-4}
    & Telematics systems & Basic; GPS tracking, fuel level, maintenance alerts, simple diagnostics & Advanced; real-time transmission and processing of detailed vehicle data (performance monitoring, predictive maintenance, remote diagnostics, and so on) \\ \cline{2-4}
    & Data collection and analytics & Limited data collection and basic analytics; track vehicle usage, maintenance schedules, and fuel consumption & Extensive data collection and advanced analytics to monitor the vehicle's performance, identify potential issues, optimize driving behavior, and enhance operational efficiency  \\ \cline{2-4}
    & Infrastructure services & Predefined data exchange with the infrastructure through static use cases.  & Extended use cases (1) \textit{Parking services}: finding, reserving, and paying for parking spaces. (2) \textit{Charging services}: locating, accessing, and paying for electric vehicle charging stations. (3) \textit{Tolling services}: automatic payment of tolls on roads or bridges, managed by integrated vehicle software.\\
    \hline
    \multirow{3}{2.1cm}{Autonomous driving} & Level of automation & Lower levels of automation (up to Level 2), relying heavily on hardware-based systems and driver intervention & Higher level of automation (Level 3 and above), with software handling complex decision-making and control  \\ \cline{2-4}
    & Hardware and sensor complexity & Hardware resources and sensors setup designed to focus on specific functions & Powerful hardware resources and sensing capabilities (e.g. high-resolution cameras) to support evolution along vehicle’s lifecycle. \\ \cline{2-4}
    & Processing architecture and AI algorithms & Decentralized processing with simpler algorithms. Focus on individual sensor data and basic automation. & Centralized processing with powerful AI algorithms, integrating combined data from different sensors to make real-time decisions \\
    \hline
    \multirow{2}{2.1cm}{Infotainment and user experience} & User interface and interaction & Limited; physical buttons and knobs, voice commands (in some models), and, in more recent models, touchscreens & Gesture controls or augmented reality displays, enhancing the overall user experience and reducing driver distraction \\ \cline{2-4}
    & Personalization & Limited; manual adjustments and preset options & Extensive; customized driving preferences and vehicle settings through software. \\
    \hline
    \multirow{4}{2.1cm}{Vehicle diagnostics and maintenance} & Diagnostics capabilities & Diagnostics rely on On-Board Diagnostic (OBD) systems. It provides limited information about the vehicle's health, such as identifying specific fault codes related to engine or emissions systems & SDVs, with advanced sensors and OBD systems, can gather extensive data on vehicle performance, including sensor data, battery health, powertrain status, and autonomous driving system performance. \\ \cline{2-4}
    & Maintenance & Hardware-based maintenance requiring regular physical checks and updates & Primarily software-based maintenance with OTA updates, maintenance scheduled proactively, and optimized vehicle uptime \\ \cline{2-4}
    & Real-time monitoring & Limited; basic warning lights or indicators on the dashboard for critical issues, such as engine malfunctions or low fuel levels & Extensive; Real-time monitoring of multiple vehicle systems and parameters allows continuous data analysis, enabling early detection of issues and prompt warnings to drivers and fleet managers. \\ \cline{2-4}
    & Predictive maintenance & Basic; often rely on preset schedules and manual inspections & Advanced predictive maintenance using data analytics and machine learning algorithms allow SDVs to predict failures and schedule maintenance proactively \\
    \hline
\end{tabular}
 \caption{Feature comparison between SDVs and conventional vehicles}
 \label{table:sdv_non-sdv} 
\end{center}
\end{table*}

\subsection{SDVs versus non-SDVs} \label{sdv_vs_non_sdv}
Key distinctions between SDVs and non-SDVs encompass control mechanisms, flexibility, upgradability, innovation potential, and maintenance approaches. Table \ref{table:sdv_non-sdv} outlines these differences in detail, emphasizing their contrasting capabilities and functionalities from a feature-oriented perspective.

 In non-SDVs, data processing is typically localized within function-specific ECUs, each dedicated to a single subsystem (e.g., braking, steering, infotainment), leading to a highly distributed architecture with limited cross-domain coordination. In contrast, SDVs adopt zonal ECUs that consolidate and manage multiple sensor and actuator interfaces within a physical zone of the vehicle (e.g., front, rear), regardless of function. This zonal organization reduces wiring complexity and enhances data aggregation efficiency. For instance, in a braking scenario, the zonal ECU collects and pre-processes data from nearby sensors, then transmits it via high-speed links (e.g., Automotive Ethernet) to the central HPC, which performs advanced computations to determine braking force. The resulting command is sent back to the zonal ECU, which translates it into real-time actuator signals. This architecture enables centralized decision-making while preserving local responsiveness and modularity.

\begin{figure}[h]
    \centering
    \includegraphics[width=0.8\linewidth]{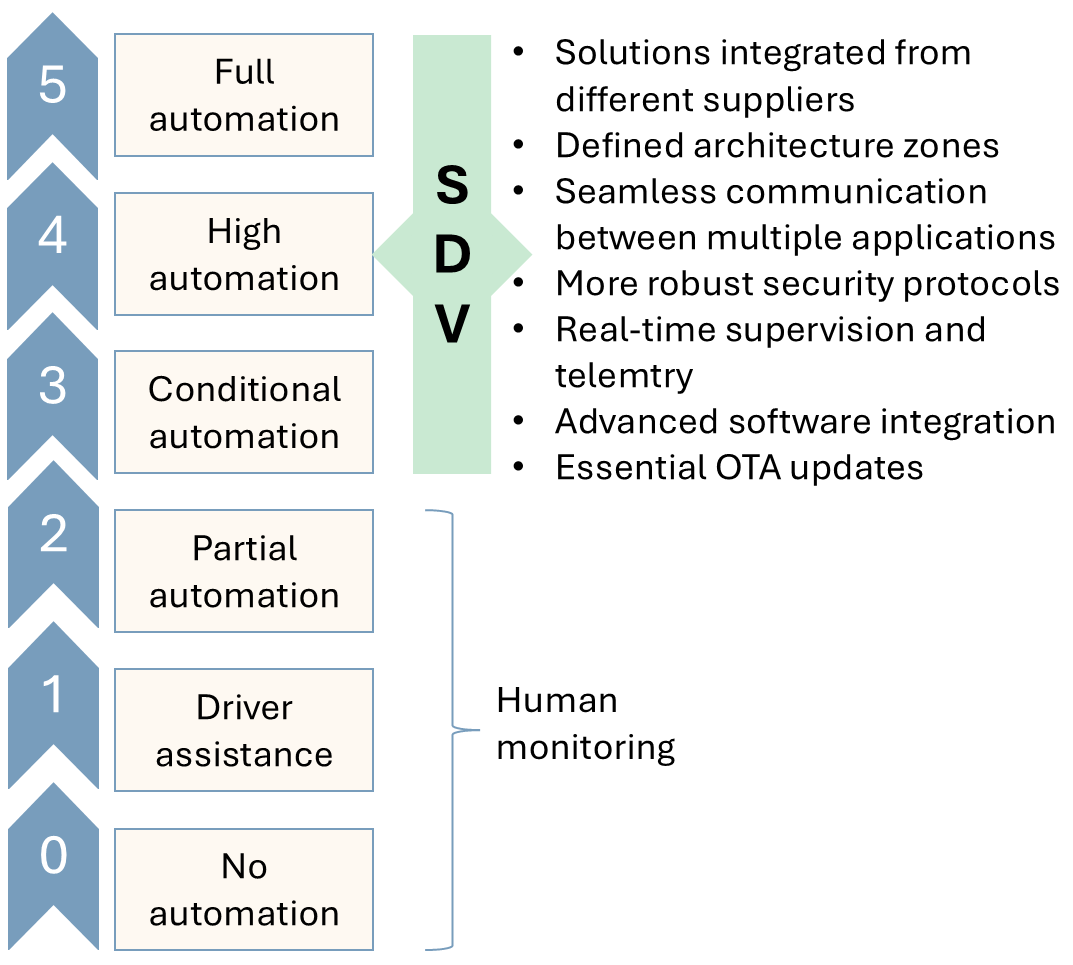}
    \caption{SAE levels of automation applicable to SDVs}
    \label{fig:automation_levels}
\end{figure}

Automation levels defined by the SAE J3016 standard \cite{SAE} play a pivotal role in SDVs. Levels 3 and above are significant due to their reliance on advanced sensors and sophisticated software for data fusion, enabling features like object detection and autonomous decision-making. These levels also support integration and communication between applications and services from various suppliers, facilitating SDV evolution. Figure \ref{fig:automation_levels} shows the concerned levels of automation for the SDV highlighting its foundational elements and key enablers. 

\subsection{SDV enabling technologies}

\subsubsection{Central computing}
The increasing demands of connectivity, electrification, shared mobility, and autonomous driving have pushed traditional E/E architectures to their limits \cite{Deloitte}. As shown in Table \ref{tab:architecture_sdv_non_sdv}, SDV architectures address these challenges by integrating an HPC as the central computing backbone \cite{continental}. The HPC decouples software from hardware, allowing continuous OTA updates for security patches, bug fixes, and new features, enhancing performance and safety throughout the vehicle's lifecycle \cite{cesa}.

Operating on POSIX-compliant Operating Systems (OS), the HPC supports cross-platform application integration, with the vehicle's OS managing and coordinating system functions for seamless performance \cite{Towards_a_RISC-V_Open_Platform_for_Next-generation_Automotive_ECUs}. This architecture enables advanced ADAS capabilities such as lane-keeping, adaptive cruise control, and emergency braking by processing real-time data from LiDAR, RADAR,  cameras, and other sensors \cite{Bosch}, delivering a robust and adaptive SDV system.
\begin{table*}[h!]
\footnotesize
\begin{center}
\begin{tabular}{|L{2.1cm}||C{6.9cm}|C{6.9cm}|}
    \hline  
        Architecture  & Non-SDV & SDV \\
        \hline \hline
        ECU &  Important number of ECUs ($\approx$ 100)  & Fewer number of Zonal ECUs \\ 
            &  Each ECU is responsible for one function &  Each zonal ECU can aggregate many functions\\
            & ECUs are connected to a central gateway through CAN buses & Zonal ECUs are connected to an HPC \\  
            & Never updated after the initial software flashing (in the manufacturing process) & Continuously updated not only for quality reasons but also on-demand to add new features \cite{Bosch} \\
        \hline
        Communication protocol &  CAN with limited bandwidth and data transmission rate (5 Mbps)  & Automotive Ethernet (AE) : provides high throughput (up to 10 Gbps) and ensures efficient and reliable communication, enabling software components to share information and collaborate with the HPC\\
        \hline
        Overall architecture &  Decentralized: ECUs interconnected with each other through CAN communication. & Zonal: presence of HPC that gathers data from zonal ECUs via AE in order to process it and take decisions \\
        \hline
\end{tabular}
\caption{Architectures comparison between SDVs and conventional vehicles}
\label{tab:architecture_sdv_non_sdv}
\end{center}
\end{table*}
\subsubsection{AI algorithms}
\begin{table*}[h!]
\footnotesize
\begin{center}
\begin{tabular}{|L{2.1cm}||C{2.1cm}|C{12cm}|}
        \hline
        Aspect & Key area & Explanation \\
        \hline
        \hline
        ADAS & Perception & AI algorithms process data from sensors like cameras, LiDAR, RADAR, and ultrasonic sensors to perceive the environment. This includes object detection, lane keeping, pedestrian detection, and traffic sign perception \cite{aptiv} \\
        \cline{2-3}
        & Decision making & AI helps in making real-time decisions by analyzing the perceived data. This includes path planning, obstacle avoidance, and vehicle control functions like acceleration, braking, and steering \\
        \cline{2-3}
        & Localization and mapping & AI algorithms, such as Deep Learning-based Simultaneous Localization and Mapping (SLAM) \cite{SLAM} help vehicles understand their precise location and create detailed maps of their surroundings\\
        \hline   
        Predictive maintenance & Fault detection & Machine learning models analyze vehicle data to predict potential failures before they happen, reducing downtime \\
        \cline{2-3}
        & Diagnostics &  AI helps in determining the optimal maintenance schedules by analyzing service intervals of  various components, thereby improving vehicle reliability and extending its lifespan \\
        \hline
        Personalization and user experience & Driver profiles & AI can create personalized driving experiences by learning and adapting to individual driving styles and preferences \\
        \cline{2-3}
        & Voice and gesture recognition & AI-powered voice assistants and gesture control systems allow for more intuitive and safer interaction with the vehicle's infotainment and control systems \\
        \hline
        Connectivity and V2X communication & Traffic management & AI enables vehicles to communicate with each other and with infrastructure (such as traffic lights) efficiently,  improving traffic flow and reducing congestion \\
        \cline{2-3}
        & Enhanced safety & V2X communication, powered by AI, allows for real-time sharing of information about road conditions, accidents, and other hazards \\
        \hline
        Cybersecurity & Threat detection & AI assists in identifying and mitigating cybersecurity threats by analyzing patterns and detecting anomalies within both the vehicle's network traffic and the broader ecosystem \\
        \cline{2-3}
        & Adaptive security measures & AI enables the implementation of adaptive security protocols that can respond in real-time to emerging threats\\
        \hline
\end{tabular}
 \caption{AI impact on SDV technology}
    \label{tab:AI_algos}
\end{center}
\end{table*}
AI algorithms are pivotal to the technological advancements of SDVs, enhancing vehicle control, user experience, and safety. Table \ref{tab:AI_algos} highlights AI's impact on advanced SDV functions. For example, \textit{Tesla} vehicles \cite{tesla_ai} use \textit{Birds-eye-view networks} to generate a top-down view of the road layout for ADAS features, leveraging neural networks on video data from multiple vehicle cameras to create a comprehensive view with 3D objects and pedestrians. 

AI algorithms also demand high processing power, typically provided by GPUs, which can be costly. In SDVs, AI models can run on the central HPC, offering the necessary resources for each function. Additionally, SDVs are designed to adapt to the evolving nature of AI models, efficiently incorporating new data and improving their performance.

\subsubsection{OTA updates}

The increasing connectivity of SDVs simultaneously heightens the need for continuous functional evolution and amplifies their exposure to cyber threats. OTA updates therefore serve not only to deliver security patches but also to distribute new features, performance improvements and configuration changes throughout the vehicle lifecycle. While traditional FOTA practices are frequently confined to subsystems such as infotainment \cite{bb_ota}, SDVs demand a flexible, cloud-based update architecture that is scalable and secure. This approach enables seamless integration of advanced features, including autonomous driving, real-time fault detection, preventive maintenance, and energy optimization based on user driving patterns. In practice, Tier 1 suppliers develop the software updates and provide them to the OEMs, who then integrate and upload these updates to secure cloud servers \cite{telematics_wire}. Vehicles then connect to these servers, via Internet, receive the latest software enhancements and security patches seamlessly and efficiently throughout their lifecycle.
\subsubsection{V2X communication}
V2X communication refers to the vehicle’s connectivity with its entire surroundings and external environment, including other vehicles (V2V), the infrastructure through the Road Side Units (RSU) (V2I), the cloud for OTA updates (V2C), the charging grid for the electric vehicles, among others.
In most current vehicles, information sharing is still limited, typically focused on basic applications like traffic signal information and hazard alerts. However, in SDVs, this capability is far more advanced and sophisticated, enabling more complex use cases such as automatic payment at electric charging station \cite{payment}, road recommendation based on traffic data shared by vehicles, and more.
Table \ref{tab:V2X_sdv_non_sdv} further summarizes the relevant differences regarding the V2X communication capabilities between SDVs and non-SDVs.
    
\begin{table*}[htb]
\footnotesize
\begin{center}
\begin{tabular}{|L{2.1cm}||C{7.1cm}|C{7.1cm}|}   
\hline
        Feature & V2X in non-SDVs & V2X in SDVs \\
        \hline
        \hline
        Integration & Limited; V2X is an add-on offered by external communication board limiting its use to few specific applications (e.g. hazard alerts, traffic signal information)  & Advanced; V2X technology is embedded in the vehicle's software architecture allowing seamless interaction with various vehicle systems (e.g. ADAS, autonomous driving modules, and infotainment systems) \\
        \hline
        Components synergy  & V2X systems function either as standalone units or are integrated  with limited aspects to the vehicle’s electronics, such as navigation, without direct communication with the vehicle's core functions & SDVs leverage advanced computing power and sophisticated software to process V2X data more effectively, enabling more complex applications, such as cooperative driving and real-time traffic optimization\\
        \hline
        Updates & Statically configured at the manufacturing phase. Managing updates is limited with this setup and it is difficult to scale to larger demands. &  The V2X system is the main gateway between the SDV and the remote management system for OTA updates. Using V2X communication technologies such as the multiple radio bearer and link aggregation, the programming of OTA updates can be intelligent, optimized, and synchronized with the driver's data \\  
        \hline
        Communication range & Current V2X implementations mainly facilitate V2V and V2I communication for straightforward tasks like collision avoidance & Broader range of applications shifting from "informative" to "integrated", offering features for an extensible vehicle \\
        \hline
  
\end{tabular}
\caption{Comparison of V2X communication capabilities in SDVs and conventional vehicles}
 \label{tab:V2X_sdv_non_sdv}
\end{center}
\end{table*}

\begin{figure*}[t!]
    \centering
    \includegraphics[width=0.8\linewidth]{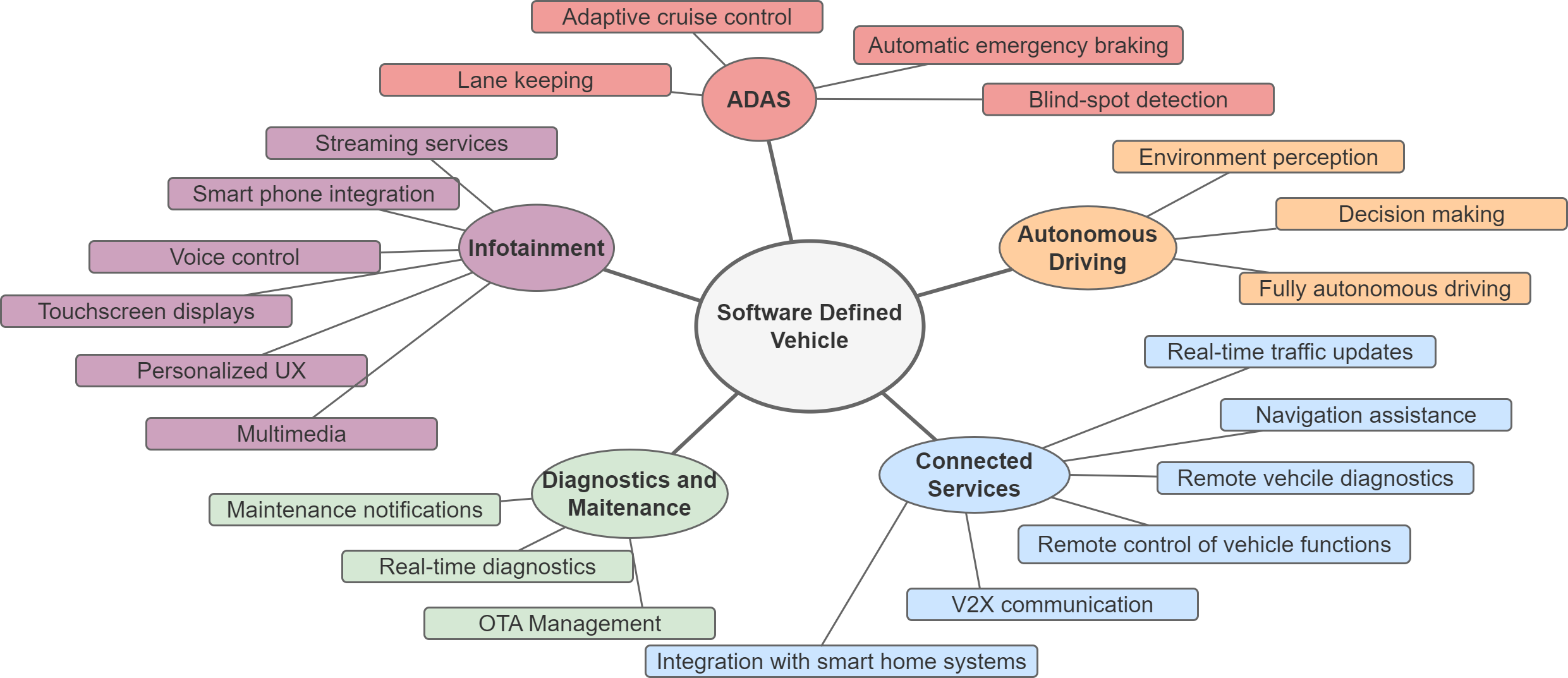}
    \caption{SDV applications grouped by functional domains}
    \label{fig:sdv_apps}
\end{figure*}

The aforementioned enabling technologies empower SDVs to host a broad spectrum of applications. Figure \ref{fig:sdv_apps} enumerates these applications and groups them by functional domain.

As SDVs become more prevalent, their complexity introduces new security challenges. The following section provides a security analysis of the SDV ecosystem, exploring vulnerabilities, threat vectors, and a comprehensive taxonomy that systematically classifies SDVs-exclusive risks across hardware, software, and communication layers, while mapping attack surfaces to their potential impacts on safety, privacy, and operational integrity.

\section{SDVs specific attack surface} \label{sec:security}

In SDVs, security has become a mandatory requirement. New regulations \cite{ISO_SAE_21434_2021}\cite{UNECE_R155_2021} mandate the need for rigorous measures regarding software management, version control, configuration management and tracking, documentation, and OTA update processes \cite{regulations}. 
An SDV can hold, manage, and synchronize different tiers' solutions and services. Hence, these regulations apply not only to OEMs, but also to suppliers because the risks in the supply chain have also increased significantly \cite{hammi2023security}.

In this section, we first define the threat model that underpins our analysis, then present the proposed taxonomy for categorizing security vulnerabilities and attack vectors in SDVs.

\subsection{Threat model}

We introduce a layered threat model and analyze the relationship of the attacks with each layer. These layers form the structural dimensions of our taxonomy.

\subsubsection{Attack flow}

Represents the pathway an attack takes from its initial exploitation to its impact on the vehicle's operations. It distinguishes between digital and hybrid attack surfaces: 
\begin{itemize}
	\item \textit{Digital attack flow:} focuses on purely software-based and cloud-driven attacks that exploit SDV connectivity, cloud services, or software dependencies. 
	\item  \textit{Hybrid attack flow:} involves both hardware and software components. Such attacks often initiate via physical access points (for examples, OBD-II ports), wireless interfaces (for examples, Bluetooth, C-V2X), or protocol vulnerabilities, and then spread through digital channels within the vehicle’s internal network such as CAN bus, AE, or LIN bus. By leveraging this dual-vector approach, adversaries can bypass isolation mechanisms and gain deeper control over critical vehicular functions.
	   \end{itemize}
	The physical attack flow is deliberately excluded from our taxonomy, as our focus is specifically on the SDV-specific attack surface, whereas purely physical attacks can affect any conventional vehicle and are thus beyond the scope of this study.
	
By classifying attacks according to their propagation flow, this dimension clarifies where vulnerabilities originate and how compromises spread, thereby aiding the identification of entry points.

\subsubsection{Target components} 

This layer highlights the SDV ecosystem elements that are exposed to cyber threats. These components can be categorized into three main areas: 
\begin{itemize}
\item \textit{Cloud infrastructure and applications:} includes cloud-hosted services, update mechanisms, training pipelines, and applications' marketplaces. 
\item \textit{Vehicle systems:} covers embedded software, communication protocols, and physical access points, which are crucial for SDV operation. 
\item \textit{Third-party ecosystems:} encompasses external entities such as suppliers and service providers, which introduce risks through software dependencies and supply chain vulnerabilities.
\end{itemize}

Understanding the target components allows us to pinpoint the most critical assets and the security measures required to protect them. 

\subsubsection{Entry points}

This layer represents the initial vector of exploitation within the SDV ecosystem. These are the interfaces or mechanisms through which an attack can infiltrate the system, including: 
\begin{itemize}
\item \textit{Update mechanisms:} e.g., OTA updates, software patching, and bug fixing. 
\item \textit{Cloud interfaces:} e.g., API vulnerabilities, and data access points. 
\item \textit{Training pipelines:} e.g., AI model poisoning, and machine/federated learning corruption.
\item \textit{Communication interfaces:} e.g., V2X, and in-vehicle networks. 
\item \textit{Supply chain delivery:} e.g., compromised third-party software. 
\end{itemize}

\subsubsection{Impacted services} 
Once an attack exploits an entry point, it affects specific services within the SDVs functionalities. These services can be classified intro four categories:
\begin{itemize}
\item \textit{Operational functionality:} attacks that disrupt in-vehicle operations, such as the communication between ECUs or QoS exploitation in V2X communication.
\item \textit{Data integrity:} threats that compromise the accuracy and trustworthiness of transmitted or stored data, such as tampered OTA update. 
\item \textit{Safety-critical systems:} the most severe attacks are those that endanger the vehicle's safety and compromise its critical functionalities. These attacks can manifest in adversarial attacks that can mislead the autonomous driving systems or zonal ECU compromises that enable remote control of critical vehicle operations, posing significant risks to passengers and road users. 
\item \textit{Privacy:} SDVs continuously exchange data with OEMs, and any breach in this ecosystem can expose sensitive user information, including location history and driving behavior. Such vulnerabilities make SDVs prime targets for surveillance, unauthorized tracking, and identity theft. 
\end{itemize}

\subsubsection{Attack vectors}

Identifies the exploitation methods used in attacks, which map directly to the SDVs properties P1–P6 previously detailed and analyzed in subsection \ref{sdv_def}.

\subsection{Proposed taxonomy}
\label{taxonomy}

Based on the defined threat model detailing adversarial capabilities, attack vectors, and system vulnerabilities in SDVs, we present a structured taxonomy to systematically organize these elements. This taxonomy not only maps the landscape of SDVs-specific threats, highlighting unique attack surfaces and patterns, but also underpins the development of precise, layer‑aligned defenses

Our taxonomy, shown in Figure \ref{fig:taxo}, focuses on SDVs-specific attacks.\footnote{While SDVs inherit many cybersecurity vulnerabilities from traditional vehicles, this taxonomy concentrates on threats that are unique to their software-centric architecture and novel interfaces.} The taxonomy hierarchically organizes attack vectors based on target components, entry points, impact domains, and timing, offering a structured approach to threat analysis and mitigation planning. The inclusion of the attack vector dimension extends the threat model by capturing the specific pathways through which adversaries exploit system vulnerabilities. This granularity enables a more precise classification of attacks, facilitating the identification of appropriate security controls for each threat scenario.
It also helps in identifying vulnerabilities and analyzing how attacks propagate through various components of SDVs. 

\begin{figure}[H]
    \centering
    \includegraphics[width=1\linewidth]{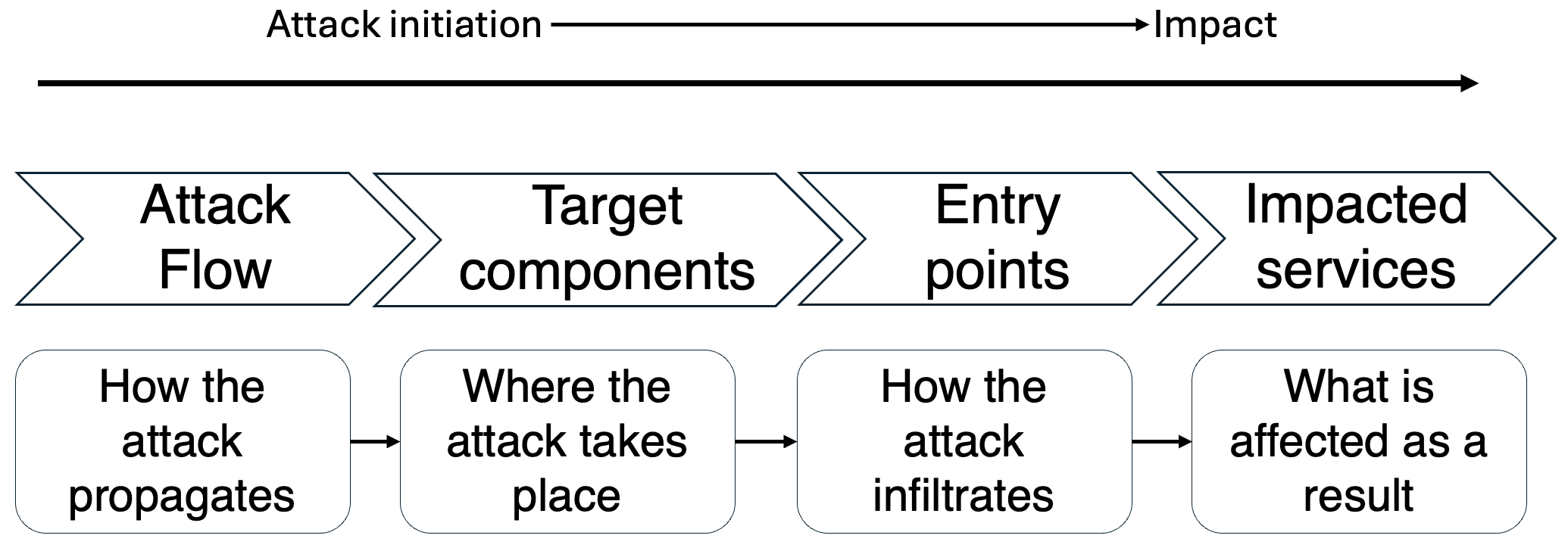}
    \caption{Logical sequence followed in the construction of the taxonomy}
    \label{fig:taxonomy_sequence}
\end{figure}

\begin{figure*}[h]
    \centering
    \includegraphics[width=1.07\linewidth]{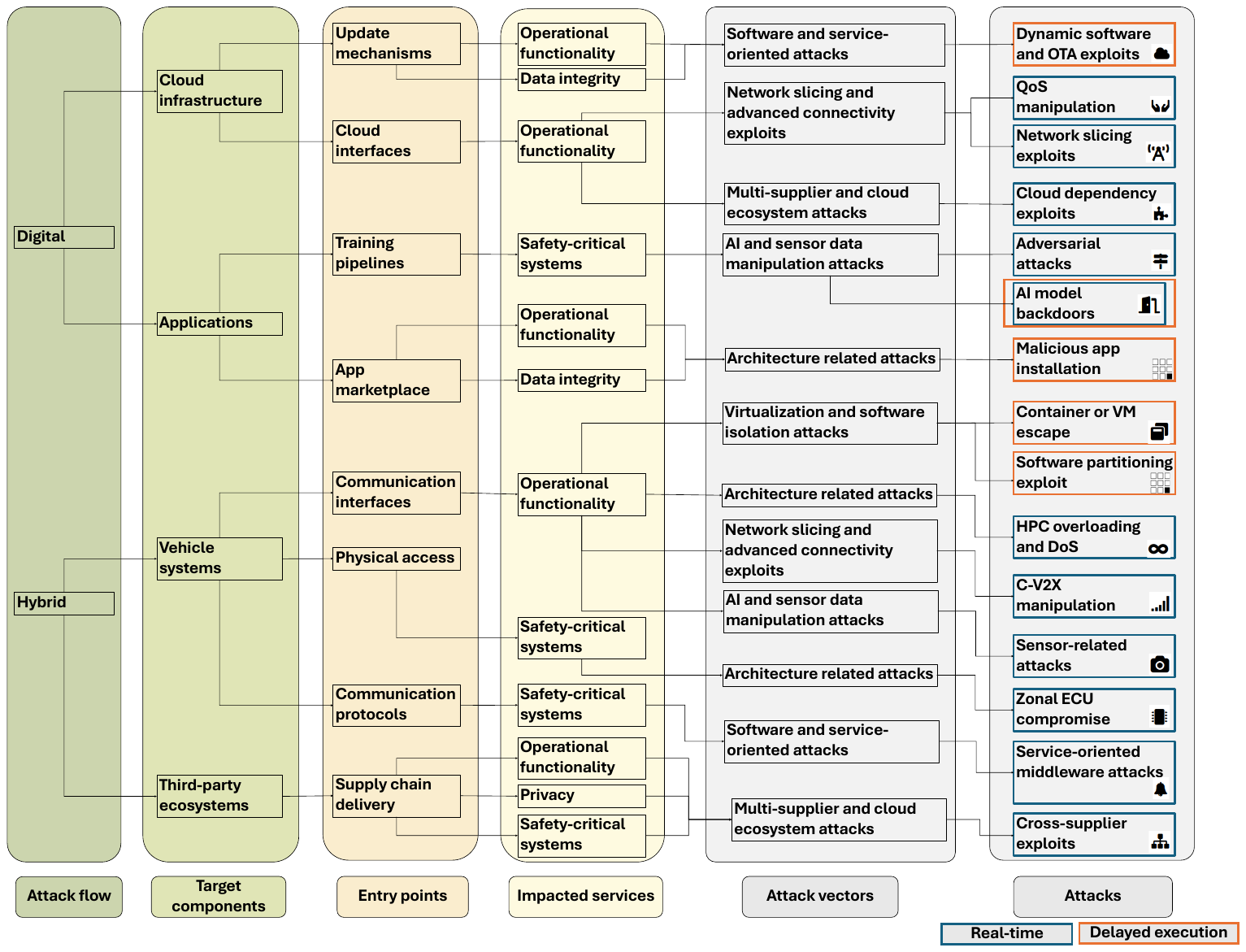}
    \caption{Proposed taxonomy of SDV-exclusive attacks}
    \label{fig:taxo}
\end{figure*}

Figure \ref{fig:taxonomy_sequence} presents a structured representation of the SDV attack lifecycle, progressing from the attack initiation to its impact. This sequential organization reflects a causality chain: the attack flow determines the propagation path, which leads to the compromise of specific components, accessed through distinct entry points, ultimately resulting in the disruption of services. Such structuring enables a holistic view of SDVs-specific threats, facilitating both the classification and the design of targeted defense mechanisms.

To enhance the characterization of cyber threats in SDVs, we introduce a classification based on their execution timing: real-time and delayed-execution attacks. This temporal distinction informs the design of tailored security protocols. Real-time attacks, such as Denial-of-Service (DoS) that immediately disrupt communication, require prompt detection and mitigation. In contrast, delayed-execution attacks, like backdoor-injected AI models that remain dormant until triggered by specific inputs, necessitate continuous integrity checks and behavioral monitoring. Differentiating these timing profiles enables the development of security mechanisms that prioritize response according to the attack’s urgency and potential for preemptive containment.

This taxonomy of SDVs-specific threats is organized according to the system layers they compromise. We begin with attacks exploiting advanced connectivity mechanisms, including network slicing and C-V2X interfaces, as they represent the primary external entry points into the SDV. We then examine vulnerabilities rooted in the vehicle's internal architecture, focusing on zonal computing and centralized processing. 
Building upon this foundation, we address threats to software services and application layers, particularly those involving partitioned environments, AI-based functionalities, and sensor-data pipelines. Finally, the taxonomy concludes with attacks targeting the broader SDV ecosystem, namely cloud infrastructures and multi-supplier integration, reflecting the extended and distributed nature of the SDVs threat surface. 
This progression reflects an external-to-internal analysis: we move from communication interfaces through the vehicle’s internal systems to broader external dependencies.

\subsubsection{Network slicing and advanced connectivity exploits}
\leavevmode\par
\textbf{Network slicing exploits \faBroadcastTower:}
network slicing and advanced connectivity exploits target vulnerabilities in the virtualized and segmented communication infrastructure of 5G-enabled SDVs. Network slicing divides a physical network into multiple virtual slices, each dedicated to specific services with tailored performance and security parameters \cite{net_slicing}.  
Attackers can exploit inter-slice interference to compromise other slices due to shared physical resources \cite{5G}. Such exploits may involve unauthorized access to high-priority slices, such as those handling safety-critical V2X communications. Attackers can also use DoS attacks (among other attacks) to degrade the performance of lower-priority, non-critical slices. 
These attacks undermine the reliability and isolation of network slices, potentially leading to data breaches, disrupted connectivity, and compromised safety especially in SDVs, where robust communication is essential for autonomous and connected functionalities. 

\textbf{QoS manipulation \faHands:}
Quality of Service (QoS) manipulation attacks exploit vulnerabilities in mechanisms that prioritize and allocate network resources to ensure reliable performance for critical applications. In the context of SDVs, QoS is essential for maintaining seamless communication, such as V2X interactions, infotainment, and safety-critical systems. 
An attacker can manipulate QoS settings by altering priority levels, bandwidth allocations, or latency parameters to disrupt the performance of high-priority services. For example, \textit{Yang et al.} in \cite{deqos} prove through their DeQoS method, that constraining channel resources of RSUs is possible to deceive the RSU into serving legitimate vehicles. 
By repeating this process with multiple vehicles, the attacker creates dummy connections that consume the RSU’s limited bandwidth and service capacity. This attack not only wastes resources, but also prevents legitimate vehicles within the RSU's range from accessing services. 

\textbf{C-V2X manipulation \faSignal:}
these attacks exploit vulnerabilities in C-V2X protocols or implementations to manipulate the integrity, authenticity, or availability of transmitted messages \cite{C_V2X_sec}. An attacker could inject false information, such as misleading location or mobility data, to disrupt traffic flow or cause accidents \cite{Fooling_LiDAR_Perception_via_Adversarial_Trajectory_Perturbation}.  
Another possible message-injection attack is the Sybil attack \cite{sybil_badis}, in which a malicious node forges multiple identities to create numerous untrusted virtual nodes, thereby undermining the assumption that received messages accurately reflect the real traffic situation. 
Finally, other tactics include targeted sidelink jamming to disrupt communication by exploiting the periodicity of Basic Safety Messages (BSMs) in C-V2X, and sidelink resource exhaustion, where the attacker strategically uses the  Semi-Persistent Scheduling (SPS) algorithm to create congestion, reducing resource availability for legitimate vehicles \cite{C-V2X_arxiv}. It's worth noting that such manipulation isn't exclusive to SDV systems, however, the dependence on C-V2X systems that rely on accurate and timely communication, such as collision avoidance, traffic management, and autonomous navigation in SDVs makes the vulnerabilities' impact bigger and more dangerous. \\

\subsubsection{Architecture related attacks}
\leavevmode\par
\textbf{Zonal ECU compromise \faMicrochip:}
the zonal ECU compromise is a significant security threat in SDVs. Zonal ECUs serve as central hubs within each zone of the vehicle, managing data flows and controlling sensors and actuators. 
A compromise of a zonal ECU enables attackers to disrupt localized functionality, such as manipulating braking, steering, or lighting systems within the compromised zone. Additionally, due to the interconnected nature of zonal architectures, an attacker can propagate malicious payloads to other zones or domain controllers, potentially causing system-wide disruptions. In the literature, various studies \cite{article14}\cite{ECU_attack} have highlighted attacks on ECUs through physical access, such as using the OBD-II interface to exploit the software system and compromise the challenge-response authentication protocol. Additionally, unauthorized hardware has been employed to inject malicious packets into the vehicle. With the advent of zonal ECUs, which oversee a wide range of critical functions, the potential impact of such attacks is significantly amplified with the architectural evolution. 

\textbf{HPC overloading and DoS \faInfinity:} 
HPC systems, with their computational power and storage capacity, face significant vulnerabilities due to their complex architecture and critical use cases. For instance, \cite{HPC_DoS} identified a wide attack surface, including weak password policies, unpatched systems, unnecessary services, and a lack of intrusion detection and prevention systems. Key vulnerabilities, including brute-force attacks on authentication mechanisms, exploitation of default or poorly configured services, and insufficient monitoring of file integrity and system logs, can be exploited to launch DoS attacks to disrupt critical HPC operations. Additionally, \cite{HPC_overloading} demonstrate that while initial flow-control checks can mitigate DoS and replay attacks at the network perimeter, these attacks may still propagate through switches to HPC and zonal ECUs, ultimately causing resource overload. 
These critical attacks can remain undetected, especially when they mimic expected high load (e.g., low and slow behavior DoS, or application-level exhaustion) or when vehicle telemetry is sparse, making malicious load hard to distinguish from normal operation, as high-performance systems are inherently designed for intensive computation and resource usage. 

These risks are further exacerbated by the expanded attack surface of HPC systems in SDVs, as operating systems running above the hypervisor layer facilitate application execution within virtualized environments \cite{HPC_containers}.  This architectural approach, while providing flexibility and resource isolation, introduces additional vulnerability points at both the hypervisor and virtual environment levels. 
As discussed earlier, in SDVs, the HPC acts as the central computing platform responsible for executing diverse software services, including safety-critical and user-facing applications. Its support for third-party application integration, akin to smartphone ecosystems, enables a wide range of functionalities but also exposes the system to new security threats. 
Specifically, the ability for users to install third-party applications creates a risk of malicious software, disguised as legitimate apps \cite{malicious_app}, exploiting vulnerabilities in the virtualized environment. 
Such exploits may lead to unauthorized access, compromise of system integrity, or interference with the safety-critical  functions of the vehicle, thereby threatening the overall security and reliability of SDVs \cite{malicious_app}.

\subsubsection{Software and service-oriented attacks}
\leavevmode\par
\textbf{Dynamic software and OTA exploits \faCloud:} 
OTA updates, while essential for maintaining and enhancing system functionality, are susceptible to various exploits that threaten their security and reliability. 
A primary vulnerability arises from weak authentication mechanisms, which may fail to verify the identities of the server or client properly, making the update process susceptible to man-in-the-middle attacks or unauthorized access. 
Similarly, the use of outdated encryption algorithms or improper storage/management of cryptographic keys, leading to insufficient protection of data and cryptographic assets, can allow attackers to intercept or manipulate updates, potentially injecting malicious software. 
Another critical risk stems from inadequately designed updates that may lack mechanisms to ensure the integrity and authenticity of update files or fail to incorporate rollback options for addressing update failures, potentially leading to system instability or complete device failure \cite{An_adaptable_security_by_design_approach_for_ensuring_a_secure_Over_the_Air_OTA_update_in_modern_vehicles}. 
Moreover, the absence of resilient recovery mechanisms, including automated rollback procedures and secure failsafe modes increases the risk of leaving systems either vulnerable or non-operational in the event of an update failure \cite{Centralized_Management_of_BLE_BasedOTA_Firmware_Updates_Design_and_Prototype_Implementation}. 

An additional and increasingly critical threat is distributed denial-of-service (DDoS) attacks targeting OTA servers. DDoS attacks can overwhelm OTA servers by flooding them with massive amounts of illegitimate requests, rendering the servers unable to respond to legitimate update requests. Such attacks not only disrupt the timely delivery of updates but also compromise the safety and security of SDVs. 
OTA update processes also share common vulnerabilities with other connected devices like smartphones and computers, including exposure to phishing attacks \cite{ota_phishing},  exploitation of unpatched software flaws, and interception of update files over unsecured networks.

\textbf{Service-Oriented Middelware over IP (SOME/IP) attacks \faBell:}
in addition to the vulnerabilities inherent to OTA update mechanisms, security concerns arise at the level of in-vehicle communication protocols. In particular, Service-Oriented Middleware over IP (SOME/IP), which serves as the backbone for service-oriented communication and dynamic interaction between ECUs, inherits many of the same risks. 

Recent studies, such as \cite{SOMEIPDOS}\cite{Analyzing_and_Securing_SOME_IP_Automotive_Services_with_Formal_and_Practical_Methods}, have demonstrated that SOME/IP is particularly vulnerable to MiTM-based strategies, including the copycat attack and the de-association attack, both of which exploit weaknesses in the service discovery and association mechanisms to redirect or terminate legitimate communications.
In the copycat attack, an adversary intercepts a legitimate  \texttt{OFFER SERVICE} message from the server and forwards a duplicate to the client, causing the client to connect to the attacker’s service and redirecting subsequent communications. 
By contrast, the de-association attack operates differently: the attacker spoofs the server to send a \texttt{STOP OFFER SERVICE} message to the client, disrupting the ongoing service. The attacker then transmits a new \texttt{OFFER SERVICE} message, coercing the client to connect to their malicious service instead. 
\texttt{OFFER SERVICE} and \texttt{STOP OFFER SERVICE} are part of the service discovery mechanism that enables communication between applications in a service-oriented architecture. These messages help clients discover available services on the network and manage their usage. 
In both scenarios, the legitimate server remains operational, and the client receives conflicting messages from both the attacker and the server. 

A more severe threat is the DoS attack, that fully disables the server’s ability to provide services or interrupts the client’s access to those services. \textit{Lehira et al.} \cite{SOMEIPDOS} have demonstrated the feasibility of such a DoS attack on real device, highlighting its potential to disrupt service provision entirely, thereby escalating the threat landscape for SOME/IP-enabled architectures. 

\subsubsection{Virtualization and software isolation attacks}
\leavevmode\par
\textbf{Container and virtual environment escape \faWindowRestore:}
SDVs predominantly use containerization and specific virtualized environments, such as automotive-grade hypervisors \cite{Cybersecurity_in_Software_Defined_Vehicles_Exposing_the_Gaps_Mapping_the_Risks}\cite{HPC_containers} and lightweight virtual electronic control units (vECUs)\footnote{\url{http://luxoft.com/blog/virtualization-revolutionizing-software-defined-vehicles-development}}. 
Container technologies like Docker and automotive-optimized systems (e.g., VirtIO\footnote{\url{http://automotive.panasonic.com/en/innovation/sdv}}) allow for isolation of vehicle functions, ensuring separation between safety-critical software (e.g., braking) and non-safety applications (e.g., infotainment), while providing resource efficiency \cite{Cybersecurity_in_Software_Defined_Vehicles_Exposing_the_Gaps_Mapping_the_Risks}.

Container/virtual environment escape is a critical security threat that occurs when an attacker exploits misconfigurations, container-runtime flaws, or kernel vulnerabilities \cite{VM_escape} to break isolation and transit from a guest instance to the host. Once escaped, the adversary can access the host OS, move laterally to other containers or VMs, exfiltrate sensitive data, disrupt services, or compromise critical vehicle functions. 

\textbf{Software partitioning exploits \faBuromobelexperte:}
software partitioning exploits target the isolation mechanisms designed to separate software components running on the same system, particularly in critical environments like SDVs. Software partitioning is intended to ensure that faults, vulnerabilities, or malicious actions in one partition do not affect others. However, attackers can exploit weaknesses in these mechanisms, such as improper enforcement of memory boundaries, insecure inter-partition communication, or misconfigured access controls. For instance, \cite{SW_partition} shows that by leveraging these vulnerabilities, an attacker can breach partition boundaries, gaining unauthorized access to restricted resources or influencing other partitions. This can lead to data leakage, manipulation of critical system operations, or the spread of malicious code across partitions, compromising the integrity and safety of the entire system. 

Software-partitioning exploits target flaws in logical isolation between software components to execute lateral movements or escalate privileges within the same host environment, whereas a container escape breaks the underlying isolation boundary (hypervisor or kernel) to move from a guest instance into the host or other guests.
In SDVs, these exploits pose a significant risk to system integrity and vehicle safety.

\subsubsection{AI and sensor data manipulation attacks}
\leavevmode\par
\textbf{Adversarial attacks \faMapSigns:}
adversarial attacks \cite{SoK_AI} can target perception systems, such as object detection or lane-keeping algorithms, leading to critical errors like misclassifying traffic signs, failing to detect obstacles, or making incorrect navigation decisions. These attacks highlight the inherent risks of relying on ML-based systems for safety-critical tasks, as they can compromise the vehicle's operational integrity and endanger passengers and other road users. For instance, \textit{Kumar et al.} \cite{adv_attack} explain how adversarial attacks target deep learning models for scene perception. They also highlight that in black-box settings, where the model's internal details are unknown, adversaries rely on confidence scores or transferability from surrogate models to craft perturbations. They also show how attack methods such as M-SimBA further exacerbate this risk by efficiently confusing the neural network, reducing its confidence in the correct class while amplifying misclassification probabilities. 

\textbf{Sensor-related attacks \faCamera:}
Sensor-related attacks \cite{SoK_AI} target the integrity, availability, or functionality of the sensors used in SDVs. \textit{El-Rewini et al.} \cite{sensor_based} have addressed different concerns in vehicular sensors' cybersecurity. On the one hand, they focused on environment sensors, such as LiDAR, cameras, radar, and GPS, that face threats like spoofing, where false signals deceive sensors into misinterpreting their environment, and jamming, which disrupts signal reception entirely. For instance, spoofing attacks on LiDAR systems can create phantom objects, while jamming GPS signals can obscure a vehicle’s location, leading to hazardous outcomes \cite{Fooling_LiDAR_Perception_via_Adversarial_Trajectory_Perturbation}. On the other hand, the authors studied vehicle dynamics sensors, including inertial sensors and Tire Pressure Monitoring Systems (TPMS), which are susceptible to physical tampering and spoofing. For example, spoofing TPMS signals can trigger false low-pressure warnings, causing unnecessary stops or even accidents. 

\textbf{AI model backdoors \faDoorOpen:}
backdoor attacks in AI models exploit weaknesses in the training process, typically by manipulating training data or injecting specific input patterns, known as triggers, that cause the model to behave maliciously under targeted conditions. Such attacks can remain dormant under normal conditions, making them particularly difficult to detect and mitigate in deployed AI systems. 

Adversarial actors exploit generative models such as \textit{CycleGAN} and \textit{Stable Diffusion}  to systematically craft poisoned training datasets embedded with specific stylistic triggers and hidden backdoors \cite{AI_back1}. These sophisticated attacks enable malicious manipulation of model predictions when particular stylistic patterns or triggers are encountered during inference, circumventing traditional detection mechanisms while maintaining model performance on clean data. The stealthy nature of such generative backdoor attacks allows adversaries to compromise machine learning systems without requiring direct access to target datasets or training processes, representing a significant threat to model integrity and trustworthiness \cite{AI_back1}.

Similarly, in object detection, attackers modify training annotations by adding trigger patterns and altering bounding boxes, causing the model to missdetect targeted objects \cite{AI_back2}. These attacks remain effective due to their stealthy implementation and resistance to conventional defenses like fine-tuning and model pruning \cite{backdoor_attack_def}.  They demonstrate how AI models can be compromised during training, leading to persistent vulnerabilities in real-world applications.

\subsubsection{Multi-supplier and cloud ecosystem attacks}
\leavevmode\par

\textbf{Cloud dependency exploits \faPuzzlePiece:}
cloud dependency exploits target the heavy reliance of SDVs on cloud-based services for functionalities such as data storage, real-time analytics, OTA updates, and V2X communication \cite{cloud_v2x}. For example, an attacker could disrupt V2X communication by targeting cloud-based traffic management services. These attacks exploit vulnerabilities in cloud infrastructure, APIs, or communication links between the vehicle and the cloud services. \cite{vehic_cloud} highlights that by compromising these dependencies, attackers can intercept or manipulate data, disrupt critical services, spoof the driver's identity, and most dangerously bypass all the security principles (authentication, confidentiality, integrity and availability). Such exploits undermine the reliability and security of SDVs, impacting their operational integrity, safety-critical functions, and user data privacy. 

\textbf{Cross-supplier exploits \faSitemap}
cross-supplier exploits take advantage of the interconnected nature of SDVs, which rely on components and services from multiple suppliers across various tiers. These exploits occur when vulnerabilities in one supplier's system or component are leveraged to compromise another supplier's system that interacts with it. For example, a vulnerability in a third-party telematics module could be exploited to gain access to the vehicle's infotainment system or even its safety-critical domain. Attackers may exploit poor integration practices, lack of standardized security protocols, or insufficient communication between suppliers to propagate their attack \cite{hammi2023security}. Such exploits highlight the risks posed by inconsistent security measures across the supply chain, potentially allowing attackers to bypass defenses and compromise multiple interconnected systems within an SDV. 

Close related are supply chain attacks. By compromising a trusted supplier or introducing malicious code into software, firmware, or components, attackers can embed backdoors or exploitable vulnerabilities that remain undetected until deployment \cite{hammi2023software}. These attacks amplify the challenges of maintaining security in a multi-tiered and highly interconnected SDV ecosystem.

Building upon the proposed taxonomy, the following section examines existing studies on cybersecurity SDVs, leveraging this structured framework to identify research gaps, highlight unresolved challenges, and contextualize current mitigation strategies within the broader threat landscape.

\begin{table*}[t!]
\footnotesize
\centering 
    \caption{Overview of key vulnerabilities and attack vectors identified in  SDV literature.\\ 
    \faCar: in-vehicle attack; \faIcon[solid]{sign-out-alt}: External/remote attack; \faDesktop: simulation; \faCogs: testbed-based evaluation, \XSolidBrush: not mentioned in the paper. }
    \label{tab:summary}
\begin{center}
\begin{tabular}{|L{0.9cm}||C{0.6cm}|C{1.9cm}|C{4.8cm}|C{1.9cm}|C{4.1cm}|C{1.0cm}|}   
\hline
        \textbf{Reference} & \textbf{Year} & \textbf{Vulnerability class} & \textbf{Attacks explored} & \textbf{Experimentation type} & \textbf{Experimentation setup} & \textbf{Attack path} \\
        \hline
        \hline
        \cite{Remote_Exploitation_of_an_Unaltered_Passenger_Vehicle} & 2015 & \faMicrochip \ \faSignal & Remote control of vehicle functions, command injection & \faCogs & Jeep Cherokee; Uconnect cellular/Wi-Fi interface & \faIcon[solid]{sign-out-alt}\\        
        \hline
        \cite{Free_Fall_Hacking_Tesla_from_Wireless_to_CAN_Bus} & 2017 & \faMicrochip \ \faSignal& Remote compromise leading to CAN injection and Remote Code Execution across ECUs & \faCogs & Tesla Model S/X & \faIcon[solid]{sign-out-alt} \\
        \hline
        \cite{Illusion_and_dazzle_Adversarial_optical_channel_exploits_against_lidars_for_automotive_applications} & 2017 & \faCamera & LiDAR perception tampering (object injection / removal) & \faCogs &  LiDAR + laser/relay setups to create spoofed point-cloud echoes  & \faIcon[solid]{sign-out-alt}\\
        \hline
        \cite{Robust_physical_world_attacks_on_deep_learning_visual_classification} & 2018 & \faCamera & Physical adversarial (stop-sign perturbations) causing misclassification & \faCogs & Printed stickers/perturbations applied to traffic signs & \faIcon[solid]{sign-out-alt} \\
        \hline
        \cite{ECU_attack} & 2018 & \faMicrochip & Buffer overflow, brute force, software manipulation & \XSolidBrush & \XSolidBrush & \faCar \\ 
        \hline
        \cite{vehic_cloud} & 2018 & \faPuzzlePiece & Data interception and manipulation & \XSolidBrush & \XSolidBrush & \faIcon[solid]{sign-out-alt} \\
        \hline
        \cite{deqos} & 2019 & \faHands & Resources overloading & \faDesktop & MATLAB &  \faIcon[solid]{sign-out-alt} \\
        \hline
        \cite{malicious_app} & 2019 & \faInfinity   & Malware injection & \XSolidBrush & \XSolidBrush & \faIcon[solid]{sign-out-alt} \\
        \hline
        \cite{Adversarial_sensor_attack_on_lidar_based_perception_in_autonomous_driving} & 2019 & \faCamera & LiDAR perception tampering (object injection / removal) &  \faCogs &  LiDAR + laser/relay setups to create spoofed point-cloud echoes  & \faIcon[solid]{sign-out-alt}\\
        \hline
        \cite{Simple_physical_adversarial_examples_against_end_to_end_autonomous_driving_models} & 2019 & \faCamera & Physical adversarial perturbations causing steering errors in end-to-end models & \faDesktop & CARLA & \faIcon[solid]{sign-out-alt} \\
        \hline
        \cite{Physgan_Generating_physical_world_resilient_adversarial_examples_for_autonomous_driving} & 2020 & \faCamera & Physical adversarial perturbations for roadside signs and objects & \faDesktop &  NVIDIA Dave-2\footnote{\url{developer.nvidia.com/blog/deep-learning-self-driving-cars/}} + Udacity dataset\footnote{https://medium.com/udacity/challenge-2-using-deep-learning-to-predict-steering-angles-f42004a36ff3}, Dave-2 testing dataset \cite{Virtual_to_real_reinforcement_learning_for_autonomous_driving}, and Kitti \cite{Vision_meets_robotics_The_kitti_dataset}  & \faIcon[solid]{sign-out-alt} \\
        \hline
        \cite{Towards_robust_LiDAR_based_perception_in_autonomous_driving_General_black_box_adversarial_sensor_attack_and_countermeasures} & 2020 & \faCamera &  LiDAR perception tampering (object injection / removal) & \faDesktop & Kitti & \faIcon[solid]{sign-out-alt} \\
        \hline
        \cite{adv_attack} & 2020 & \faMapSigns & Adversarial attacks & \faDesktop & GTSRB dataset \cite{data1} & \faIcon[solid]{sign-out-alt} \\
        \hline
        \cite{sensor_based} & 2020 & \faCamera & Spoofing, jamming, physical tampering & \XSolidBrush & \XSolidBrush & \faIcon[solid]{sign-out-alt} \\
        \hline
        \cite{article14} & 2021 & \faMicrochip & Malware injection & \XSolidBrush & \XSolidBrush & \faCar \\
        \hline
        \cite{C_V2X_sec} & 2022 & \faSignal & Sidelink jamming and resource exhaustion & \faCogs & Cohda MK6C V2X onboard units, Ettus USRP B210 \cite{usrp} &  \faIcon[solid]{sign-out-alt} \\
        \hline
        \cite{sybil_badis} & 2022 & \faPuzzlePiece & Sybil attack & \faDesktop & Koln dataset \cite{Generation_and_analysis_of_a_large_scale_urban_vehicular_mobility_dataset} & \faIcon[solid]{sign-out-alt} \\
        \hline 
        \cite{DDoS_OTA} & 2023 & \faCloud  & DDoS & \faDesktop & CPLab Simulator & \faIcon[solid]{sign-out-alt} \\ 
        \hline            
        \cite{mmspoof_Resilient_spoofing_of_automotive_millimeter_wave_radars_using_reflec_array} & 2023 & \faCamera & Radar spoofing & \faCogs & Reflect-array spoofer, off-the-shelf automotive radar units & \faIcon[solid]{sign-out-alt} \\
        \hline
        \cite{HPC_overloading} & 2023 & \faInfinity  & Flooding & \faCogs & Modified production vehicle & \faCar \\
        \hline
        \cite{AI_back1} & 2023 &  \faDoorOpen & Data poisoning & \faDesktop & CIFAR10 \cite{data2} dataset & \faIcon[solid]{sign-out-alt} \\
        \hline
        \cite{AI_back2} & 2023 &  \faDoorOpen & Data poisoning & \faDesktop & ImageNet \cite{data3} datasets & \faIcon[solid]{sign-out-alt} \\
        \hline
        \cite{Revisiting_lidar_spoofing_attack_capabilities_against_object_detection_Improvements_measurement_and_new_attack} & 2023 & \faCamera & LiDAR perception tampering (object injection / removal) &  \faCogs & LiDAR + laser/relay setups to create spoofed point-cloud echoes & \faIcon[solid]{sign-out-alt}\\
        \hline
        \cite{SOMEIPDOS} & 2024 & \faBell  & Copycat, de-association, DoS & \faCogs & Real device; laptops, Open vSwitch, vsomeip & \faIcon[solid]{sign-out-alt} \\
        \hline
        \cite{CAN_MIRGU_A_Comprehensive_CAN_Bus_Attack_Dataset_from_Moving_Vehicles_for_Intrusion_Detection_System_Evaluation} & 2024 & \faMicrochip  & Physical CAN injection: DoS, spoofing, replay, and fuzzing & \faDesktop & Dataset collected from a modern vehicle (specific model not disclosed) & \faCar \\
        \hline
        \cite{CICIoV2024} & 2024 & \faMicrochip  & Spoofing, DoS & \faCogs & 2019 Ford car & \faCar \\
        \hline        
        \cite{Cyberattack_Resilience_of_Autonomous_Vehicle_Sensor_Systems_Evaluating_RGB_vs_Dynamic_Vision_Sensors_in_CARLA} & 2025 & \faCamera &  LiDAR perception tampering, GPS spoofing, DoS  & \faDesktop & CARLA & \faIcon[solid]{sign-out-alt}\\
        \hline         
        
\end{tabular}
\end{center}
\end{table*}

\section{Discussion} 
\label{sec:summary}

Table \ref{tab:summary} summarizes our analysis of existing studies that demonstrate attacks on vehicle infrastructures. Given the relative scarcity of works explicitly addressing SDVs, we broadened the scope to include research targeting autonomous and next-generation connected vehicles whose operational functions overlap with SDVs capabilities. However, because attacks on vehicular networks, particularly V2X-originated vectors, have been extensively surveyed  \cite{Security_issues_and_challenges_in_V2X_A_survey}\cite{A_survey_of_security_and_privacy_issues_in_v2x_communication_systems}, generic V2X network-only studies are omitted from Table \ref{tab:summary}. We instead focus on works that demonstrate attacks specifically pertinent to SDVs architectures.
For each study we systematically map the evaluated attack onto the taxonomy of attack types presented in Figure \ref{fig:taxo}, and report the key attributes, mainly whether the attack was simulated or executed on real hardware, the precise experimental setup, and the attack path exploited during the intrusion.
The inclusion of attack paths highlights the diverse entry points that adversaries exploit, spanning in-vehicle and out of the vehicle (external/remote) access. 

From Table \ref{tab:summary} we observe that a plurality of evaluated attacks target vehicle perception systems (LiDAR, camera and radar sensors). This concentration reflects two complementary factors: (1) perception sensors are inherently exposed to the external environment and thus accessible to physical or radio frequency adversaries, and (2) they provide the raw inputs on which autonomy and safety-critical decisions depend, so small manipulations can produce large, system-level effects. Consequently, attacks against perception afford high impact with comparatively low attacker effort, highlighting the need for sensor-centric hardening and cross-modal validation in SDVs.

Although conventional cyberattacks such as spoofing, DoS and sensor manipulation remain prevalent, their consequences are significantly amplified in SDVs. This amplification arises from SDVs' heavy reliance on sensors for environmental perception and AI for autonomous driving. A single sensor-based attack can cascade into catastrophic events, underlining the criticality of securing these systems. Additionally, emerging attack vectors, such as supply chain compromises and cloud-based exploits, remain underexplored in the vehicular context. 
Although these attack surfaces remain insufficiently explored in the specific context of SDVs, existing research on HPC systems and container isolation offers valuable analogies, as these environments share comparable challenges in terms of resource management, isolation mechanisms, and security vulnerabilities.

The effectiveness of mitigation techniques is highly contingent upon both the specific experimental setups and the underlying simulation frameworks. Subtle variations in factors such as the attack vector, model architecture and parameters, dataset characteristics, and defense implementation can lead to pronounced differences in reported robustness or detection rates. Consequently, conclusions about the efficacy of a given mitigation approach are often only reproducible within narrowly defined evaluation settings.
Among the analyzed references, only \cite{HPC_overloading} explored a zonal architecture by manually modifying the internal network of a production vehicle. 
This isolated effort highlights a notable research gap: a substantial number of attacks have yet to be evaluated under conditions specific to SDVs environments, primarily due to their emerging and evolving nature. The software-centric architecture of SDVs introduces a distinct set of vulnerabilities, ranging from software bugs and compromised machine learning pipelines to cloud-based threat vectors, that are either absent or significantly less prominent in conventional automotive systems. Furthermore, the OTA propagation of such vulnerabilities adds an additional layer of complexity, amplifying the challenges associated with securing SDVs' infrastructures.

Traditional vehicular networks, are commonly simulated using frameworks like OMNET++, SUMO, and Veins. However, these tools are limited in addressing the modularity and software-dependency intrinsic to SDVs \cite{Platelet_Pioneering_Security_and_Privacy_Compliant_Simulation_for_Intelligent_Transportation_Systems_and_V2X}. 
From Table \ref{tab:summary} we observe that no single simulation environment or framework has emerged as a community reference. Instead, authors predominantly rely on bespoke toolchains built using public datasets and custom extensions. This heterogeneity undermines reproducibility and comparability. Indeed, as for experimental setups, results are sensitive to dataset biases, sensor models, and implementation details. 
Hence, subtle differences in attack parameters or evaluation metrics can materially change outcomes. The reliance on ad-hoc setups also limits realism (many simulations omit end-to-end stack or timing effects) and impedes systematic benchmarking of defenses. To address these drawbacks the field needs shared benchmark suites, standardized sensor and platform models, harmonized metrics and more open testbeds that enable repeatable, comparable evaluations across studies.
Future research should prioritize the advancement and integration of sophisticated simulation methodologies, including digital twins \cite{digital_twin} and cloud-based emulators, due to their capacity to more accurately represent the dynamic, heterogeneous, and highly interconnected nature of SDVs. Digital twins, by creating real-time, high-fidelity virtual replicas of physical systems, enable continuous monitoring and predictive analysis of vehicle behavior under varying operational scenarios. Cloud-based emulators complement this by providing scalable and flexible platforms capable of simulating large-scale network interactions and diverse attack scenarios that are infeasible to reproduce in physical testbeds. Together, these methodologies overcome the inherent limitations of conventional simulation and testing by enabling comprehensive end-to-end evaluation of SDVs software stacks, communication protocols, and security mechanisms in environments that closely mimic real-world conditions.

The evolving supply chain of SDVs, characterized by a shift toward open-source software \cite{eclipse_sdv} and collaboration among diverse suppliers, presents both opportunities and challenges. A promising direction is the development of collaborative frameworks that facilitate cross-organizational sharing of SDV simulation tools. Such initiatives can accelerate research, standardize testing methodologies, and foster innovation while mitigating risks associated with fragmented development efforts.

The inherent complexity of the SDV ecosystem, coupled with its extensive attack surface, reinforces the necessity for proactive security approaches. Reactive measures alone are insufficient to address the rapidly evolving threat landscape. Notably, most documented SDV attacks are executed remotely, as Table \ref{tab:summary} shows, emphasizing the openness and interconnectedness of the SDV ecosystem. This underscores the urgent need for robust, system-wide security strategies that address both the digital and hybrid dimensions of SDVs, along with enhanced mitigation techniques adapted to the expanded attack surface.

\section{Challenges and open issues}
\label{sec:challenges}
\begin{figure}[h!]
    \centering
    \includegraphics[width=1\linewidth]{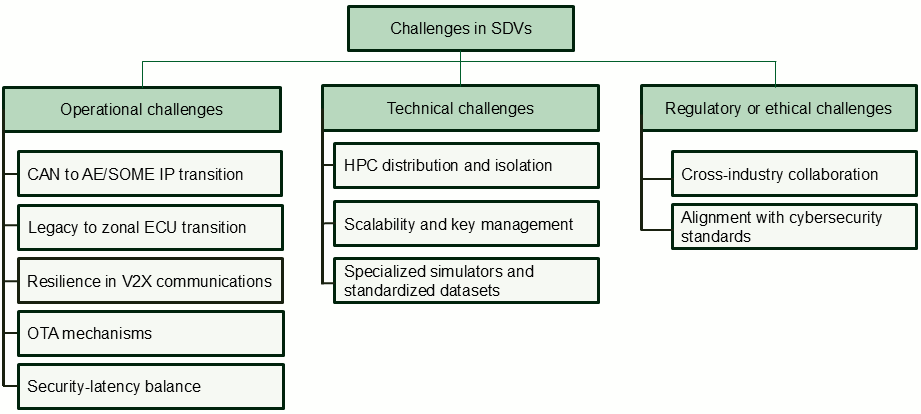}
    \caption{Taxonomy of the security challenges facing SDVs}
    \label{fig:taxo_challenges}
\end{figure}

The increasing complexity of SDVs introduces numerous security oriented challenges that must be addressed to ensure the safety, reliability, and resilience of these systems. Figure \ref{fig:taxo_challenges} summarizes these challenges  classified based on their technical, operational, and regulatory dimensions, providing a structured overview of the key security concerns in SDV systems.

\subsection{Technical challenges:}

One major technical and architectural challenge in SDVs concerns the deployment and functional distribution of multiple HPCs rather than relying on a single centralized unit. 
The rationale for deploying multiple HPCs stems from the increasing computational demands and the need for redundancy and fault tolerance inherent to safety-critical vehicular applications. Distributing HPCs allows for workload partitioning, improved real-time responsiveness, and containment of failures within isolated computing domains, thereby enhancing overall system resilience. 
However, this architectural choice raises important questions regarding the segregation of safety-critical and non-safety-critical functionalities across different HPCs. Such separation is imperative to reduce attack surfaces and prevent a single compromised HPC from propagating vulnerabilities or failures throughout the entire vehicle system. Concurrently, virtualization technologies, particularly the use of virtualized environments and containers, are employed to enable flexible allocation and isolation of software components within and across HPCs. While these containers facilitate modularity and efficient resource management, their security hinges on the robustness of their isolation mechanisms. 
Weaknesses in hypervisors or inter-container communication channels can be exploited to bypass isolation boundaries, thereby undermining system security. 
Therefore, ensuring strong and verifiable container isolation is critical to maintaining the integrity and confidentiality of both safety-critical and non-safety-critical functions distributed over multiple HPCs.

The scalability of the security infrastructure and the update management systems is critical as the number of SDVs grows. Secure key management and fleet management become increasingly complex, requiring solutions that can handle a large volume of vehicles without compromising security. The challenge here is balancing the need for scalable infrastructure with personalized updates and security. For example, large scale fleet management systems must securely handle OTA updates across thousands of vehicles while ensuring that even vehicles requiring specialized updates receive the correct software tailored to their unique operational context and configuration. Key management solutions must be robust enough to support dynamic, secure communication between vehicles and the infrastructure, while avoiding vulnerabilities that could arise from centralized systems.

Another notable technical challenge in advancing SDVs cybersecurity lies in the lack of specialized simulators and the absence of widely recognized and standardized datasets tailored to these systems \cite{Platelet_Pioneering_Security_and_Privacy_Compliant_Simulation_for_Intelligent_Transportation_Systems_and_V2X}. Similar to how the development of automotive control systems benefited from specific testbeds and Hardware-in-the-Loop/Model-in-the-Loop (HiL/MiL) platforms \cite{Designing_distributed_controlling_testbed_system_for_supply_chain_and_logistics_in_automotive_industry} for functional validation, the security domain now requires dedicated simulation tools that can emulate the unique communication patterns, software modularity, and attack surfaces of SDVs.

These gaps hinder the validation of mitigation strategies and the reproducibility of research, underscoring the urgent need for advanced simulation platforms and comprehensive datasets to support robust and scalable solutions.

\begin{tcolorbox}[colback=gray!10, colframe=black, boxrule=0.3pt]
\textbf{Insights:} it is challenging to apply security mechanisms like micro-segmentation into zonal architectures promoted by SDVs, with the prominence of a central HPC.
At the same time, OTA update frameworks must securely deliver both fleet-wide and personalized updates. Addressing these challenges is further hindered by the absence of standardized cybersecurity simulators and datasets tailored to SDVs environments, limiting the reproducibility and validation of proposed security solutions.
\end{tcolorbox}

\subsection{Operational challenges:}

The transition from CAN to Automotive Ethernet (AE) introduces another significant challenge. AE provides higher bandwidth but comes with increased complexity and susceptibility to attacks such as spoofing and flooding \cite{A_systematic_review_on_security_attacks_and_countermeasures_in_automotive_ethernet}. This shift necessitates the development of robust security mechanisms to handle these vulnerabilities and ensure a smooth transition that aligns with evolving standardization efforts. Furthermore, SOME/IP is designed to work seamlessly over AE. Although SOME/IP is regarded as a highly promising middleware for service-oriented architectures like SDV's, it lacks built-in security features to safeguard applications and transmitted data against malicious attacks \cite{someip_sec, SOMEIPDOS} which necessitates the inclusion of additional cybersecurity protection mechanisms. 

In terms of connectivity, the reliance on V2X communications for autonomous driving introduces significant risks \cite{article3}\cite{survey_v2x}\cite{survey2_v2x}. V2X is integral to decision-making processes in SDVs, yet it is highly susceptible to interference, jamming, and spoofing attacks. These vulnerabilities highlight the importance of developing resilient and secure communication protocols to ensure the safety of autonomous systems.

Memory and processing limitations inherent to legacy ECUs represent significant security challenges, as many of these units lack the computational capacity to implement fundamental security mechanisms such as cryptographic operarions. This deficiency exposes them to a wide range of potential attacks, undermining the overall integrity of vehicular systems. The ongoing shift toward zonal ECU architectures promises substantial improvements in processing power and memory capacity, which could facilitate the integration of advanced security functions alongside emerging demands like AI processing and large-scale data collection. Nevertheless, it remains an open question whether these enhanced resources will be adequate to simultaneously fulfill stringent security requirements and the computationally intensive tasks introduced by the next-generation of vehicular applications, highlighting the need for further investigation into resource allocation and optimization strategies within zonal ECU frameworks \cite{Power_hungry_processing_Watts_driving_the_cost_of_ai_deployment}.

OTA updates, a cornerstone of modern SDVs functionality, face their own set of challenges. Current practices, which involve reprogramming entire ECU images, are inefficient given the limited data rates available \cite{Secure_over_the_air_software_updates_in_connected_vehicles_A_survey}. The shift to delta updates, where only modified code is transmitted, introduces concerns about the security of patching algorithms and the potential latency and complexity these updates may incur. Moreover, the timing and synchronization of updates must be carefully managed, as ECUs perform distinct functions depending on the vehicle’s state. Poor synchronization could result in vulnerabilities during critical operations.

Balancing real-time safety-critical functionalities with real-time attack detection and response is another pivotal issue. Advanced anomaly detection systems and fail-safe mechanisms must be developed to mitigate attacks without introducing latency that jeopardizes operational safety. This dual focus on safety and security remains a pressing concern.

\begin{tcolorbox}[colback=gray!10, colframe=black, boxrule=0.3pt]
\textbf{Insights:} operational security can no longer be treated as a layer atop legacy systems, but must be re-architected to accommodate concurrent demands such as flexible software updates, real-time safety guarantees, and secure communication interfaces. This calls for moving beyond isolated security measures toward integrated, system-level approaches that combine safety and security the full lifecycle, from the design phase through deployment.
\end{tcolorbox}

\subsection{Regulatory challenges:} 

The influence of telecomunication operators on TCU hardware and service delivery adds another layer of complexity. Telecomunication operators play a pivotal role in maintaining the infrastructure that underpins connectivity. However, misalignment between operators and automotive cybersecurity standards could create vulnerabilities that compromise the safety and reliability of SDV systems. 
For instance, past incidents such as the BMW ConnectedDrive vulnerability have demonstrated how weak/inadequate security implementations can expose critical vehicle functions to remote exploitation \cite{BMW_ConnectedDrive_Vulnerability}\cite{cybersecurity_challenges_in_vehicular_communications}\cite{Beemer_Open_Thyself_Security_vulnerabilities_in_BMW_s_ConnectedDrive}. In this case, attackers could intercept unencrypted HTTP communications between 2.2 million BMW, Mini, and Rolls-Royce vehicles and BMW's servers by deploying a fake cellular base station (IMSI catcher) combined with 3G/4G signal jamming to force vehicles to connect via vulnerable 2G networks. The attack exploited multiple security flaws: (1) BMW used identical symmetric encryption keys across all vehicles, (2) transmitted critical commands via unencrypted HTTP GET requests formatted as XML without Transport Layer Security (TLS) protection, and (3) failed to implement server certificate verification. 
This allowed attackers to forge unlock commands, monitor vehicle locations and speed data, access private emails sent through BMW Online services, and even modify emergency contact numbers stored in the vehicle, all within minutes and without leaving digital traces.

Broader security concerns extend to the supply chain and internal communication mechanisms. With diverse OEMs and tier-one suppliers contributing to vehicle production, securing the supply chain becomes increasingly challenging. Third-party components in hardware and software introduce potential risks that could compromise the entire vehicle ecosystem, necessitating stringent vetting processes and robust security policies. 

Addressing these challenges requires a multi-faceted approach that includes robust architectural designs, secure communication protocols, and advanced update mechanisms. Furthermore, fostering collaboration among stakeholders, including OEMs, suppliers, and telecom operators, is essential to develop and implement security standards that can evolve alongside advancements in autonomous and next generation vehicle technologies.

\begin{tcolorbox}[colback=gray!10, colframe=black, boxrule=0.3pt]
\textbf{Insights:} the regulatory landscape for SDVs is strained by the convergence of automotive, telecommunications and software domains. To ensure trust and resilience, regulation must shift toward unified, cross-sector guidelines and continuous-compliance models that reflect the dynamic lifecycle of SDV development and deployment. AUTOSAR \cite{Autosar_standard} constitutes an initial technical step toward standardisation, but it primarily targets operating-system and middleware interfaces and is currently implemented by a limited set of manufacturers; therefore regulatory efforts must extend beyond to cover the full ecosystem.
\end{tcolorbox}

\subsection*{Over-The-Air updates, the primary risk}
We argue that OTA updates represent the most critical cybersecurity challenge in SDVs, given their central role in maintaining, patching, and upgrading vehicle software throughout the lifecycle. A single compromised update can jeopardize the integrity of millions of vehicles simultaneously, making OTA channels a high-value target for attackers. Furthermore, the OTA update process involves multiple stakeholders, including OEMs, Tier-1 suppliers, telecom operators, and cloud service providers, each potentially using different security protocols, and infrastructure models.

The inherent complexity of OTA update processes in SDVs highlights a critical research imperative: \textbf{the development of a secure-by-design OTA update protocol} capable of coordinating updates across heterogeneous platforms while ensuring end-to-end security and adherence to regulatory standards. Addressing this challenge necessitates initiating the design process with a thorough and systematic risk assessment, aimed at identifying relevant threat vectors and delineating attack surfaces specific to diverse stakeholders. Subsequently, the protocol must incorporate a multi-layered security framework that aligns with established industry and regulatory guidelines, including ISO/SAE 21434 \cite{ISO_SAE_21434_2021}, ISO/SAE 24089 \cite{ISO_SAE_24089_2023}, and UNECE WP.29 recommendations \cite{UNECE_WP29}, thereby guaranteeing security, robustness, and trustworthiness by design.

\section{Conclusion}
\label{sec:conclusion}
This paper underscores the critical importance of cybersecurity in SDVs and seeks to stimulate further discourse on potential threats and their corresponding countermeasures. Through a preliminary analysis, we delineate the key distinctions between SDVs and traditional vehicles, emphasizing how the extensive connectivity and software dependence of SDVs render them more susceptible to security vulnerabilities.

We systematized the SDVs-specific vulnerabilities and examined the security implications of the broader ecosystem, spanning in-vehicle components and protocols in the SDV zonal architecture, and external interactions such as supply chains and cloud infrastructures. Our analysis of existing vulnerabilities highlights a recurring pattern of similar design choices adopted by many OEMs, offering insights into shared risks. Furthermore, by examining current attack strategies in SDVs and related domains, we project potential future threats and extract critical challenges and directions for advancing SDVs cybersecurity.

By mapping these vulnerabilities, we aim to foster a comprehensive understanding and encourage the development of robust security-by-design measures to protect the integrity and safety of SDVs. 
A secure-by-design appraoch must be implemented across all phases of the vehicle lifecycle, embedding protective measures at every stage. In this context, OTA update mechanisms become essential for reinforcing and preserving the resilience and operational performance of SDVs against evolving threat landscapes.

In light of these considerations, our future research will focuse on advancing the area of security-by-design OTA update protocols, with the aim of developing a lightweight, scalable, and standards-compliant solution that addresses the complex security challenges specific to SDV environments.

\bibliographystyle{IEEEtran}
\bibliography{biblio}

\end{document}